\documentclass[12pt]{article}
\makeatletter
\newsavebox{\@brx}
\newcommand{\llangle}[1][]{\savebox{\@brx}{\(\m@th{#1\langle}\)}%
  \mathopen{\copy\@brx\kern-0.5\wd\@brx\usebox{\@brx}}}
\newcommand{\rrangle}[1][]{\savebox{\@brx}{\(\m@th{#1\rangle}\)}%
  \mathclose{\copy\@brx\kern-0.5\wd\@brx\usebox{\@brx}}}
\makeatother

\usepackage[pdftex]{graphicx}
\usepackage{array}
\usepackage{amsmath}
\usepackage{rotating}
\usepackage{longtable}
\usepackage{mathtools}
\usepackage{tikz}
\usepackage[linguistics]{forest}
\usepackage{parskip}
\usepackage{bigints}
\usepackage{adjustbox}
\usepackage[utf8]{inputenc}
\usepackage{tikz}
\usetikzlibrary{shapes.geometric,arrows}
\tikzstyle{startstop} = [rectangle, rounded corners, minimum width=3cm, minimum height=1cm,text centered, draw=black]
\tikzstyle{io} = [trapezium, trapezium left angle=70, trapezium right angle=110, minimum width=3cm, minimum height=1cm, text centered, draw=black]
\tikzstyle{process} = [rectangle, minimum width=3cm, minimum height=1cm, text centered, draw=black]
\tikzstyle{decision} = [diamond, minimum width=3cm, minimum height=1cm, text centered, draw=black]
\tikzstyle{arrow} = [thin,->,>=stealth]

\usepackage[square,numbers]{natbib}
\usepackage{floatrow}
\usepackage{subfig}

\usepackage{float}
\usepackage{array}
\usepackage{geometry}
\usepackage{color}
\usepackage{lipsum}

\usepackage{diagbox} 
\usepackage{pdfpages}
\usepackage{multicol}
\usepackage[utf8]{inputenc}
\usepackage[english]{babel}
\usepackage{mathtools}
\usepackage{bbm}
\usepackage{amssymb}
\usepackage{amsthm}
\usepackage{hyperref}
\usepackage{bm}
\usepackage{mathtools}
\usepackage{fancybox}
\usepackage{multirow}
\usepackage{authblk}
\usepackage{algorithm}
\usepackage{algpseudocode}
\usepackage{setspace}
\usepackage{caption}
\usepackage{fancybox}
\usepackage{forest}
\usepackage{tikz-qtree}
\usepackage{graphicx}
\allowdisplaybreaks
\newlength\szg
\newcommand\quan[1]{%
\settoheight\szg{#1}%
\tikz[baseline]{\pgfmathparse{
    ifthenelse(#1 < 10, 1, ifthenelse(#1 < 100, 0.75, 0.5))
}
\let\hfs\pgfmathresult
\node at (0,\szg/2) {\makebox[0em][c]{\scalebox{\hfs}[1]{#1}}};
\draw (0,\szg/2) circle (\szg/2+0.35ex);
}}

\makeatletter
\newcommand*{\rom}[1]{\expandafter\@slowromancap\romannumeral #1@}
\makeatletter
\def\BState{\State\hskip-\ALG@thistlm}
\makeatother
\usepackage{xcolor}
\renewcommand{\raggedright}{\leftskip=0pt \rightskip=0pt plus 0cm}

\title{\bf  Filtration-Based Learning of Multiscale Shared
Structures for Multiple Functional Predictors }
\author[1]{Shuhao Jiao\thanks{Corresponding Author: shuhao.jiao@cityu.edu.hk}}
\author[2]{Hernando Ombao}
\author[3]{Ian W.\ McKeague}
\affil[1]{Department of Biostatistics\\  City University of Hong Kong}
\affil[2]{Statistics Program,  KAUST, Saudi Arabia}
\affil[3]{Department of Biostatistics\\ Columbia University}
\date{}

\begin{document}
	\maketitle			
	\setlength\parindent{0pt}
	\setlength{\parskip}{1em}
	\theoremstyle{definition}
	\newtheorem{theorem}{Theorem}
	\newtheorem{lemma}{Lemma}
	\newtheorem{prop}{Proposition}
	\newtheorem{Definition}{Definition}
	\newtheorem{corollary}{Corollary}
	\newtheorem{remark}{Remark}
	\newtheorem{assumption}{Assumption}

\begin{abstract}
It is crucial to learn the shared structures among functional predictors, as these structures characterize how predictor components exert common effects and, more generally, how predictors are homogeneously associated with the response. However, learning from multiple functional predictors is challenging because response-predictor dependencies may vary across representation dimensions and emerge at multiple resolutions, ranging from globally shared effects to predictor-specific effects. To address this issue, we propose a filtration-based shared structure learning framework for multiple functional predictors. The proposed framework organizes predictors through a hierarchical forest structure, in which shared and predictor-specific components are progressively identified from coarse to fine filtration layers. Building on this structure, we develop a filtration-based pursuit  pipeline for shared structure discovery, together with a filtrated functional partial least squares method for shared component extraction and coefficient estimation under the learned shared structures. Simulation studies show that the proposed framework is able to recover the dominant coarse-to-fine organization of the underlying shared structures and yield improved prediction performance relative to competing methods. Applied to lower-limb angular kinematics, the proposed framework improves evaluation accuracy and reveals interpretable joint coordination patterns associated with aging. More broadly, it provides a new multiscale representation-learning perspective for complex data consisting of multiple multidimensional objects.\\

\end{abstract}
\newpage

\section{Introduction}
Functional regression with multiple functional predictors has found broad applicability across a wide range of scientific domains (see e.g., \cite{ref6,ref24,ref26,ref27,ref33,ref16,ref38,ref3,ref7}). 
Learning from multiple functional predictors requires modeling not only predictor--response associations but also the shared structures among predictors. In many applications, these structures evolve across multiple resolutions. 
Effective modeling therefore requires learning how predictors interact, what information is shared, and at which scales such shared structures contribute to the response. 
A representative example arises in gait analysis (see e.g., \cite{ref17,ref22,ref28}), where angular kinematic trajectories from different lower-limb joints exhibit coordinated age-related patterns. These trajectories are functionally coupled through human biomechanics. Some shared components correspond to globally coordinated motion signatures, whereas others are specific to subsets of joints or individual joints. 
Such multiscale structures are rarely known a priori, and pre-specifying them may introduce severe misspecification and substantially degrade both representation quality and predictive performance. We therefore need a data-adaptive framework to learn multiscale shared structure directly from data, rather than relying on pre-specified structures.

Grouping approaches (see e.g., \cite{ref31,ref18,ref23,ref30,ref15}) and related work on multiscale modeling (see e.g., \cite{ref39,ref45,ref46}) provide useful perspectives for exploiting shared structures of multiple functional predictors, because they explicitly represent shared structures by assigning predictors with similar response-associated effects to the same group, so that shared features can be learned within each group. 
However, these methods remain limited in their ability to capture multiscale shared structures. In particular, these approaches rely on a fixed grouping structure or do not allow for partially shared effects among subsets of predictors. This is too restrictive when the globally shared, partially shared, and predictor-specific effects coexist across different layers of representation. 
Multiscale learning is not yet well explored in multiple functional regression. 
To address this challenge, we develop a filtration-based framework that allows the data to determine which multiscale structure is strongly supported. The proposed framework represents predictors through a hierarchical forest architecture, in which shared and predictor-specific components are progressively extracted from coarse to fine filtration layers. As a result, it enables the joint modeling of globally shared, partially shared, and predictor-specific associations within a unified multiscale representation.
Our contributions are threefold:
\begin{itemize}
\item[1)] We formulate the multiscale shared-structure learning problem under dimension-varying predictor homogeneity.
\item[2)] We develop a filtrated functional partial least squares algorithm for shared component extraction and model estimation.
\item[3)] We propose a data-adaptive structure-learning procedure, consisting of forest-structured grouping and prediction-aligned structure selection, to identify shared structures across representation dimensions rather than relying on pre-specified structures.
\end{itemize}

The {\it advantage} of the proposed framework lies in its ability to capture  multiscale structures in a flexible, data-driven manner. This represents a meaningful improvement over approaches based on pre-specified structures, since such structures may be misspecified and fail to reflect the true dependency pattern in real data, as well as over approaches that impose a single structure across all layers. 
This new line of work bridges statistical modeling and structured representation learning, enabling flexible modeling of complex multiscale dependencies while preserving interpretability. More broadly, our work introduces filtration as a general principle for structured learning in multivariate functional data, opening a new direction for representation learning and model structure discovery in infinite/high-dimensional settings. 

The rest of this paper is organized as follows. Section \ref{pre} introduces the problem setup and preliminaries. Section \ref{method} presents the proposed filtration-based multiscale shared-structure learning framework, including the multiscale representation, the filtrated functional partial least squares estimation procedure, and the data-driven structure-learning algorithm. Section \ref{s4} reports the simulation settings and numerical results, with emphasis on structure recovery and prediction performance under different synthetic scenarios. Section \ref{s5} validates the proposed framework on lower-limb angular kinematics for age-related gait connectivity analysis and demonstrates its interpretability and predictive advantage in effective age evaluation. Section \ref{s6} concludes the paper and discusses directions for future research.

\section{Problem Formulation and Model Representation}
\label{pre}
We first introduce the basic setup and then reformulate the multiscale structures across representation dimensions.
Consider a scalar response $y_n$ and $p$ functional predictors $\{X_{jn}(t)\colon j=1,\ldots,p\}$. 
It is assumed that $\{y_n,X_{jn}(t)\colon j=1,\ldots,p\}$ are independent across $n=1,\ldots,N$, where $N$ is the sample size. 
Multiple functional regression with a scalar response $y_n\in\mathbb{R}$ and multiple functional predictors $\{X_{jn}(t)\in \mathcal H\triangleq L^2[0,1]\colon j=1,\ldots,p\}$ is given below, 
\begin{equation}
\label{mlr}
y_n=\beta_0+\sum_{j=1}^{p}\langle X_{jn},\beta_j\rangle+\epsilon_n,\\\ \mbox{E}\epsilon_n=0,\ \mbox{Var}(\epsilon_n)=\sigma^2,
\end{equation} 
where the coefficient functions $\{\beta_j(t)\}_{j=1}^p\in \mathcal H^p$. 
Suppose that, for some sets of basis functions $\{\nu_{jd}(t)\in \mathcal H\colon d\ge1\}$ and $\{u_{jd}(t)\in \mathcal H\colon d\ge1\}$, the functional predictors and functional coefficients admit the following basis representation
$$X_{jn}(t)=\sum_{d\ge1}\xi_{jn,d}\nu_{jd}(t),\ \beta_j(t)=\sum_{d,d'\ge1}b_{jd}M_{j,dd'}u_{jd'}(t),$$ 
in which the basis functions $\{\nu_{jd}(t)\colon d\ge1\}$ and $\{u_{jd}(t)\colon d\ge1\}$ can either be orthonormal or not. The  infinite matrix $\{M_{j,dd'}\colon d,d'\ge1\}$ satisfies the condition $$\sum_{d'\ge1}M_{j,d_1d'}(\int\nu_{jd_2}u_{jd'})=\bm{1}_{\{d_1=d_2\}}.$$ Then model \eqref{mlr} can be equivalently rewritten in the following form 
\begin{equation}
\label{basis-form}
y_n=\beta_0+\sum_{d=1}^\infty\sum_{j=1}^p \xi_{jn,d}b_{jd}+\epsilon_n,\\\ \mbox{E}\epsilon_n=0,\ \mbox{Var}(\epsilon_n)=\sigma^2. 
\end{equation} 
We focus on the representation in  \eqref{basis-form}, as it is dimension-specific and therefore naturally supports multiscale structure discovery across representation layers.

\section{Methodology}
\label{method}
\subsection{Preliminaries on Fixed Grouping Representation}

We first introduce the ordinary fixed grouping formulation because it provides a natural basis for learning shared features within each group under a uniform coefficient structure, thereby illustrating the key role of grouping in shared-structure learning and motivating a new multiscale framework.
Under fixed grouping, predictors sharing similar response-associated effects are assigned to the same group, yielding a conventional grouped regression model. 
Given a grouping structure $G=(\mathcal{K}_1,\ldots,\mathcal{K}_m)$ satisfying  $\mathcal{K}_i\cap \mathcal{K}_{i'}=\emptyset$ for $i\ne i'$ and $\cup_{i=1}^m\mathcal{K}_i=\{1,\ldots,p\}$, the ordinary grouped model is 
\begin{equation}
\label{grouped_model}
y_n=\beta_0+\sum_{i=1}^m\langle Z_{in},\alpha_i\rangle+\epsilon_n,\\\ \mbox{E}\epsilon_n=0,\ \mbox{Var}(\epsilon_n)=\sigma^2, 
\end{equation} 
where $Z_{in}(t)=\sum_{j\in \mathcal{K}_i}X_{jn}(t)$ and $\alpha_i(t)$ is the common coefficient function of group $\mathcal{K}_i$, capturing the homogeneous association shared by the predictors within the group. The unknown grouping structure $G$ is identified by optimizing the objective function with a pairwise fusion penalty $P_\lambda(\bm{\beta}(t))=\sum_{i<j}J_{\lambda}(\|\beta_i-\beta_j\|)$:
\begin{align*}
\label{loss}
S_\lambda(\bm{\beta}(t))&=\frac{1}{2}\sum_{n=1}^N\left(y_n-\sum_{j=1}^p\langle X_{jn},\beta_j\rangle\right)^2+P_\lambda(\bm{\beta}(t)),
\end{align*}
where $\bm{\beta}(t)=\{\beta_{j}(t)\colon j=1,\ldots,p\}$, and $J_{\lambda}(\cdot)$ is some concave penalty, such as the SCAD penalty \cite{ref11} and the MCP \cite{ref36}. 
As $\lambda$ increases, predictors tend to be clustered into fewer groups, yielding a sequence of grouping structures that collectively define the grouping path $\mathcal{P}$, from which one grouping structure is selected for the grouped model \eqref{grouped_model}.

The ungrouped model \eqref{mlr} ignores shared structure across predictors and becomes increasingly inefficient as the number of predictors grows, whereas the grouped model in \eqref{grouped_model} assumes a fixed grouping configuration across all representation dimensions. This assumption is too restrictive when shared structure evolves across resolutions.
Indeed, as demonstrated in Section \ref{s5}, both shared and predictor-specific components of different angular kinematic trajectories play important roles in effective age evaluation, and imposing a uniform structure across all representation dimensions results in suboptimal model performance. 
\subsection{Multiscale Representation}
\begin{figure}[tb]
\center
\includegraphics[width=10cm]{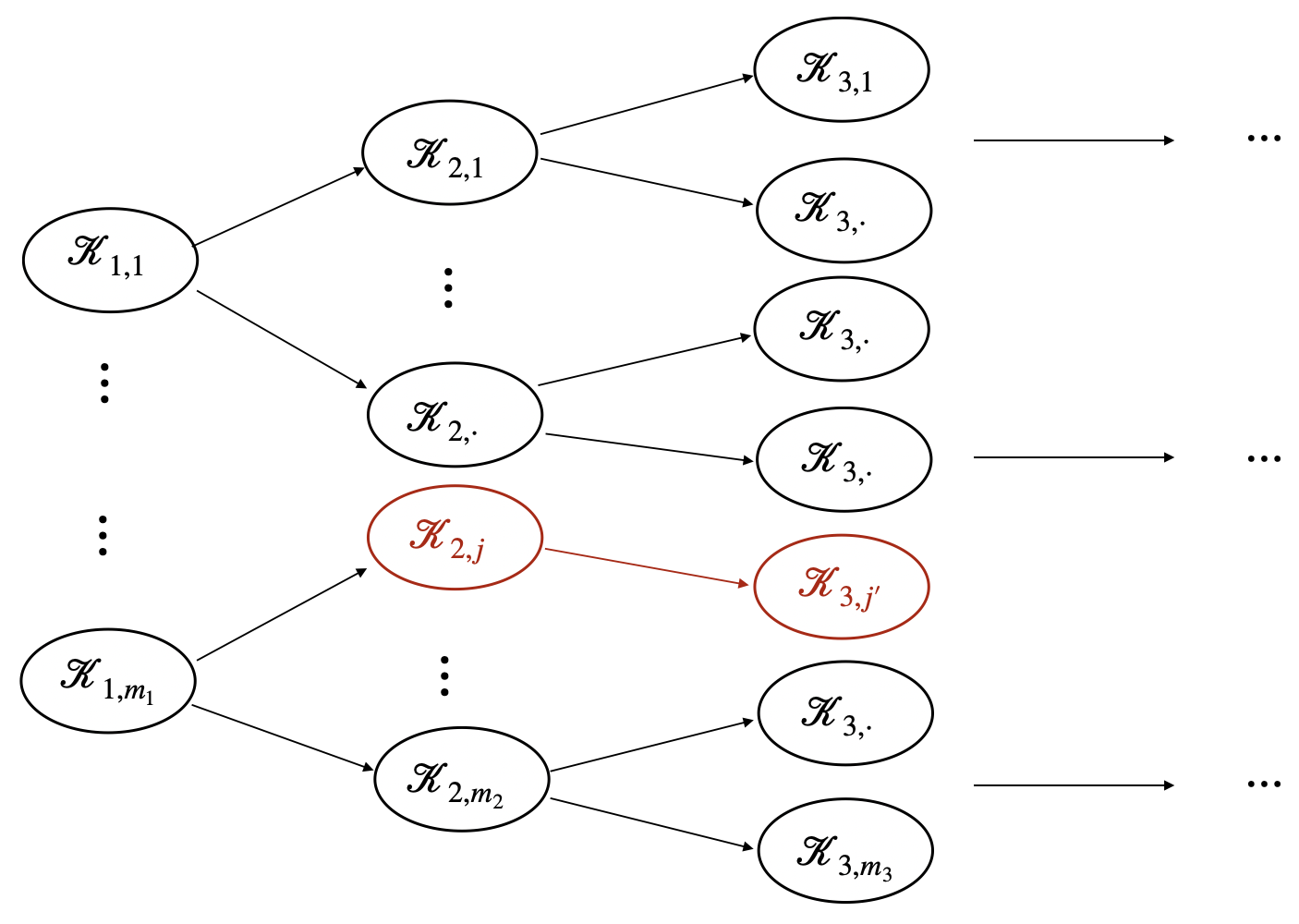}
\caption{An example of hierarchical forest structure of coefficient shared structure. Each group $\mathcal K_{d,i}, d\ge2$ is either nested within or identical to (e.g., $\mathcal K_{2,j}=\mathcal K_{3,j'}$) its ``parent" group (i.e., the group from which it originates through a split) in the previous layer. }
\label{tree}
\end{figure}

We model multiscale shared structures among functional predictors through a hierarchical forest architecture.
Each filtration layer defines a grouping configuration at a particular resolution, and progressively refines shared structure from globally shared patterns to predictor-specific effects. 
Building upon the model representation \eqref{basis-form}, we develop the following multiscale representation
\begin{equation}
\label{filt-group}
y_n=\beta_0+\sum_{d=1}^\infty\sum_{i=1}^{m_d}\left\{\sum_{j\in\mathcal{K}_{d,i}}\xi_{jn,d}\right\}a_{d,i}+\epsilon_n,
\end{equation}
where $\mathcal{K}_{d,i}$ denotes the $i$-th group in the $d$-th dimension/filtration layer and $m_d$ is the number of groups in the $d$-th dimension. For each $\mathcal K_{d,i}$, there always exists a $j$ so that $\mathcal K_{d,i}\subseteq\mathcal K_{d',j}$ for $d'<d$. Consequently, the resulting group indices $\{\mathcal{K}_{d,i}\colon i\ge1,d\ge1\}$  form a forest structure (see Figure \ref{tree}). We also require that $\mathcal{K}_{d,i}\cap \mathcal{K}_{d,i'}=\emptyset$ for $i\ne i'$ and $d\ge1$. The predictor scores $\xi_{jn,d}$ in the same group $\mathcal K_{d,i}$ share the same coefficient score $a_{d,i}$. 
In addition, to reveal the common association pattern of predictors within the same group, we project the predictors in $\mathcal K_{d,i}$ onto a common direction $\psi_{d,i}(t)$, thereby obtaining layer-specific scores for constructing the shared component, defined as
$$\zeta_{nd}^{(i)}\triangleq\sum_{j\in\mathcal{K}_{d,i}}\xi_{jn,d},\ i \geq 1,\ d \geq 1.$$
Compared with the grouped model in \eqref{grouped_model}, the multiscale model in \eqref{filt-group} allows grouping configurations to vary across representation dimensions, and deeper layers correspond to finer-resolution shared structure.
These components capture response-relevant information shared among predictors within each group, which are expected to have strong associations with the response. In particular, when $\mathcal{K}_{d,i}$ contains only a single predictor, $\zeta_{nd}^{(i)}$ captures the predictor-specific feature. In addition to obtaining these layer-wise shared components, we also need to estimate the associated coefficient scores $\{a_{d,i} \colon i \geq 1,\ d \geq 1\}$. 

Note that model \eqref{filt-group} is generally not identifiable in a strict layer-by-layer sense. To illustrate, suppose that two consecutive layers induce the same grouping configuration, and let their corresponding layer-wise shared components associated with the same group be $\zeta_{n,d}$ and $\zeta_{n,d+1}$. Their contribution to \eqref{filt-group} then takes the form
\(
a_1\zeta_{n,d}+a_2\zeta_{n,d+1}.
\)
Now define a new aggregated component 
\(
\tilde{\zeta}_{n,d}=b_1\zeta_{n,d}+b_2\zeta_{n,d+1},
\)
where $b_1,b_2\neq 0$. If we further write 
\(
a_1=b_1\tilde{a},\ a_2=b_2\tilde{a},
\)
then
\(
a_1\zeta_{n,d}+a_2\zeta_{n,d+1}
=
\tilde{a}\bigl(b_1\zeta_{n,d}+b_2\zeta_{n,d+1}\bigr)
=
\tilde{a}\,\tilde{\zeta}_{n,d}.
\)
Hence, the same contribution of shared components can be represented either by two separate layer-wise shared components or by a single aggregated component. However, this does not imply that only one layer is needed for each grouping configuration. Even under the same grouping structure, multiple layers are still necessary for fully extracting the response-relevant variation carried by those groups, since different layer-wise shared components correspond to different response-aligned directions. In this sense, the non-identifiability discussed above concerns only the exact layer-by-layer parameterization, not the practical need for multiple layers to capture the dominant response-associated structure. 

Although the proposed multiscale structures cannot fully represent non-nested shared structures, they still provide an effective response-aligned approximation by sequentially capturing the dominant shared effects across layers. In this sense, the proposed framework is not intended to exactly recover arbitrary shared structures, but rather to construct a multiscale representation that explains the response-relevant variation {\it more effectively than fixed structures and unstructured alternatives}.
Therefore, our goal is not exact recovery of a so-called ``oracle" decomposition,  but rather discovery of a statistically reasonable multiscale shared structure that captures the dominant response-relevant organization among predictors. This also helps explain why, in our simulations, the proposed method favors a more parsimonious representation with fewer layers. 


Using these multiscale grouping structures, we can simultaneously separate globally shared, partially shared, and predictor-specific association patterns at multiple resolutions across filtration layers. In the first layer, we identify the shared components with the broadest shared scope across predictors and group them accordingly to obtain $\{\zeta_{n1}^{(i)}\colon i\ge1\}$ and then estimate $\{a_{1,i}\colon i\ge1\}$. After filtering out these layer-wise shared components from each predictor, we refine $\{\mathcal{K}_{1,i}\colon i\ge1\}$ into $\{\mathcal{K}_{2,i}\colon i\ge1\}$, where each new group is either nested within or identical to a group in the previous layer. We then extract the shared components and estimate the corresponding coefficient scores in the new layer. This procedure continues until a predefined stopping criterion is met. 
As the filtration proceeds, the method progressively moves from globally shared components to partially shared and predictor-specific components. Note that if some predictors exhibit substantially different effects, so that globally shared structure is negligible, they may remain unfused with the others across all filtration layers. In the extreme case, if no shared structures are present, then the multiscale model reduces to the ungrouped model \eqref{basis-form}.

The framework is termed filtration-based because it constructs the shared components through an ordered sequence of nested layers, moving from broad shared patterns to finer predictor-specific effects. This leads to two key questions: given the forest structure, how can we extract the layer-wise shared components and estimate the coefficient scores, and how can we determine a reasonable forest structure? In Section~\ref{filt-PLS}, we address the first problem by developing a filtrated partial least squares approach. In Section~\ref{s3}, we develop a data-driven algorithm to construct the forest structure.

\subsection{Filtrated Partial Least Squares Algorithm (filt-PLS)}
\label{filt-PLS}
To extract the layer-wise shared components and estimate the coefficient scores, we develop a filtrated functional partial least squares method. 
The guiding principle is that the extracted shared components should have strong explanatory power for the response. 
The proposed filt-PLS method is designed for dimension-specific grouping configurations, under which classical functional partial least squares (fPLS, see \cite{ref8}) is no longer directly applicable.
To build intuition, we first consider a simple special case.
In the ordinary grouped model \eqref{grouped_model}, if $X_{1n}(t),\ldots,X_{pn}(t)\in \mathcal{K}_1$ (i.e., all predictors belong to a single group $\mathcal{K}_1$), then we can estimate the model using the classical fPLS approach, where $\sum_{j\in\mathcal{K}_1}X_{jn}(t)$ is treated as the predictor. The first step is to maximize $\mbox{cov}(y_n,\sum_{j\in\mathcal{K}_1}\langle X_{jn},\psi\rangle)$ over $\psi(t)$, subject to the constraint $\|\psi\|=1$. The resulting feature $\sum_{j\in\mathcal{K}_1}\langle X_{jn},\psi\rangle$ is used to explain the response, and $\psi(t)$ represents the shared direction that characterizes the common response-associated effect of the predictors in $\mathcal{K}_1$. This is followed by a projection step, and the two steps are then repeated iteratively. In the filtrated grouped model \eqref{filt-group}, however, the shared structure varies across resolutions. As a result, the maximization and projection steps are iteratively performed under different grouping configurations. 

We now detail the algorithmic pipeline. When $d=1$, we first find the filt-PLS basis by maximizing the covariance between the response and the shared component for each group, as follows, 
\begin{align*}
\psi_{1,i}(t)&\triangleq\arg\max_{\|\psi\|=1}\mbox{cov}\left\{y_n,\left\langle\sum_{j\in\mathcal{K}_{1,i}}X_{jn},\psi\right\rangle\right\}=\frac{\mbox{cov}\{y_n,\sum_{j\in \mathcal{K}_{1,i}}X_{jn}\}}{\|\mbox{cov}\{y_n,\sum_{j\in \mathcal{K}_{1,i}}X_{jn}\}\|}(t),
\end{align*}
where $i=1,\ldots,m_1$. 
The $m_1$ first-layer shared components are obtained as $\zeta_{n1}^{(i)}=\sum_{j\in \mathcal{K}_{1,i}}\int X_{jn}\psi_{1,i}$ for $i=1,\ldots,m_1$, and then estimate the associated coefficient scores: 
$$\hat{a}_{1,i}=\arg\min_{\{a_{1,i}\colon i=1,\ldots,m_1\}}\mbox{E}\left\{y_n-\sum_{i=1}^{m_1}a_{1,i}\zeta_{n1}^{(i)}\right\}^2.$$
For $d\ge2$, 
the filt-PLS basis for group $\mathcal{K}_{d,i}$ is 
\begin{align*}
\psi_{d,i}(t)&\triangleq\arg\max_{\|\psi\|=1}\mbox{cov}\left\{y_n^{[d]},\left\langle\sum_{j\in \mathcal{K}_{d,i}}X^{[d]}_{jn},\psi\right\rangle\right\}=\frac{\mbox{cov}\{y_n^{[d]},\sum_{j\in \mathcal{K}_{d,i}}X^{[d]}_{jn}\}}{\|\mbox{cov}\{y_n^{[d]},\sum_{j\in \mathcal{K}_{d,i}}X^{[d]}_{jn}\}\|}(t).
\end{align*}
We then obtain the shared components $\zeta_{nd}^{(i)}=\sum_{j\in \mathcal{K}_{d,i}}\int X^{[d]}_{jn}\psi_{d,i}$ and estimate the associated coefficient scores as follows
$$\hat{a}_{d,i}=\arg\min\limits_{\{a_{d,i}\colon i=1,\ldots,m_d\}}\mbox{E}\left\{y^{[d]}_n-\sum_{i=1}^{m_d}a_{d,i}\zeta_{nd}^{(i)}\right\}^2,$$
where 
\begin{align*}
X^{[d]}_{jn}(t)&=X^{[d-1]}_{jn}(t)-\sum_{i=1}^{m_{d-1}}\zeta_{n,d-1}^{(i)}\phi_{d-1,i}(t),\\
\phi_{d,i}(t)&= \arg\min_{\{\delta_i(t)\colon i\ge1\}}\mbox{E}\left\|X^{[d]}_{jn}(t)-\sum_{i=1}^{m_{d}}\zeta_{nd}^{(i)}\delta_i(t)\right\|^2,\\
y_n^{[d]}&=y_n^{[d-1]}-\sum_{i=1}^{m_{d-1}}\zeta_{n,d-1}^{(i)}\hat{a}_{d-1,i}.
\end{align*}
Particularly, $X_{jn}^{[1]}(t)=X_{jn}(t)$, $y_n^{[1]}=y_n$. 

In each layer, the filt‑fPLS basis functions and the corresponding coefficient scores jointly account for homogeneous association patterns of predictors in the same group. Note that under forest structure, the layer-wise shared components across different layers are orthogonal (i.e., $\sum_{n\ge1}\zeta_{nd}^{(i)}\zeta_{nd'}^{(i')}=0$ for $d\ne d'$ and $i,i'\ge1$). Consequently, the layer-wise shared components identified at different layers capture distinct information, allowing the filtrated grouped model \eqref{filt-group} to represent multi-resolution structure without redundancy and with clear interpretability. At each filtration layer, the method identifies response-aligned shared components under the current grouping configuration, removes the explained variation from both predictors and response, and then proceeds to a finer layer. In practical applications, the unknown covariances and expectations are replaced by their empirical counterparts as summarized in Algorithm \ref{al1}. 
\begin{algorithm}[!t]
\caption{Filtrated functional partial least squares}
\label{al1}
\begin{algorithmic}[1]
\State Initialize $d \gets 1$, $X_{jn}^{[1]}(t) \gets X_{jn}(t)$, and $y_n^{[1]} \gets y_n$.
\While{the stopping criterion in Section~\ref{s3.3} is not met}
    \For{$i=1,\ldots,m_d$}
        \State Estimate the filt-PLS basis:
        \[
        \hat{\psi}_{d,i}(t)=
        \frac{\sum_{n=1}^{N}\sum_{j\in\mathcal{K}_{d,i}} y_n^{[d]}X_{jn}^{[d]}(t)}
        {\left\|\sum_{n=1}^{N}\sum_{j\in\mathcal{K}_{d,i}} y_n^{[d]}X_{jn}^{[d]}\right\|}.
        \]
    \EndFor
    \State Compute the layer-wise shared components:
    \[
    \zeta_{nd}^{(i)}=\sum_{j\in\mathcal{K}_{d,i}}
    \left\langle X_{jn}^{[d]},\hat{\psi}_{d,i}\right\rangle,
    \qquad i=1,\ldots,m_d.
    \]
    \State Estimate the coefficient scores by least squares:
    \[
    (\hat{a}_{d,1},\ldots,\hat{a}_{d,m_d})
    =
    \arg\min_{\{a_{d,i}\}_{i=1}^{m_d}}
    \sum_{n=1}^{N}
    \left(
    y_n^{[d]}-\sum_{i=1}^{m_d} a_{d,i}\zeta_{nd}^{(i)}
    \right)^2.
    \]
    \State Update the predictor residuals:
    \[
    X_{jn}^{[d+1]}(t)
    =
    X_{jn}^{[d]}(t)-\sum_{i=1}^{m_d}\zeta_{nd}^{(i)}\hat{\phi}_{d,i}(t),
    \]
   where
    \[
    \hat{\phi}_{d,i}(t)
    =
    \arg\min_{\delta_i(t)}
    \sum_{n=1}^{N}
    \left\|
    X_{jn}^{[d]}(t)-\sum_{i=1}^{m_d}\zeta_{nd}^{(i)}\delta_i(t)
    \right\|^2.
    \]
    \State Update the response residuals:
    \[
    y_n^{[d+1]}
    =
    y_n^{[d]}-\sum_{i=1}^{m_d}\hat{a}_{d,i}\zeta_{nd}^{(i)}.
    \]
    \State $d \gets d+1$.
\EndWhile
\end{algorithmic}
\end{algorithm}

\subsection{Multiscale Shared Structure Learning}
\label{s3}
\subsubsection{Structure Learning Principle}
Exhaustive search over all admissible hierarchies is computationally infeasible especially when $p$ is large, so we need to first identify a statistically meaningful and computationally tractable subset of candidate structures for constructing the shared structures. Our strategy is guided by the grouping path $\mathcal{P}$. As established in the existing literature (see \cite{ref53,ref54,ref31}), the grouping path is piecewise linear, and groups with more similar response-associated effects tend to merge earlier along the path, indicating stronger empirical evidence of homogeneity in earlier fusions. In this sense, the grouping path provides a statistically guided and computationally tractable collection of candidate grouping structures for learning the unknown multiscale hierarchy. 
Unlike the ordinary grouped model \eqref{grouped_model}, which uses only a single grouping structure from the grouping path, we construct filtration layers by hierarchically organizing multiple grouping structures along the path from coarse to fine resolutions. Thus, the ordinary grouped model \eqref{grouped_model} can be viewed as a special case of the multiscale grouped model \eqref{filt-group}. Note that we do not strictly follow the grouping path; rather, we focus only on those grouping structures that retain comparatively strong explanatory power. 
In addition, we restrict attention to grouping structures that satisfy a nestedness property, whereby groupings at deeper layers refine those at earlier layers. This restriction is not merely algorithmic: nested grouping structures are essential for ensuring a coherent hierarchical decomposition of shared and predictor-specific components, which underpins the interpretability of the resulting forest structure. 
Different penalties or grouping paths may lead to different candidate hierarchies, as is common for data-driven structure-learning procedures. We therefore do not treat the grouping path as the hierarchy itself, but rather as a statistically informed source of candidate grouping configurations from which a coherent filtration structure is constructed. 
More details are given in Section \ref{selection}.

\subsubsection{Candidate Forest Generation}
\label{selection}
After obtaining the grouping path $\mathcal{P}$, we first construct candidate sets of grouping structures from the path, from which the forest structure is then constructed. Notationally, for two grouping structures $G,G'$, $G \subseteq G'$ means that every group in $G$ is nested in some group in $G'$. 
Denote a candidate set by $\mathcal{G}=\{G_{1},\ldots,G_{\ell}\}$, where $\ell$ is the number of grouping structures in $\mathcal{G}$, and $G_{i}\in\mathcal{P}$ denotes the $i$-th layer's grouping structure of the candidate set $\mathcal{G}$ satisfying $G_{\ell}\subseteq G_{\ell-1}\subseteq\cdots\subseteq G_{1}$.

Multiple candidate sets can be constructed from the grouping path, and only one of them is used for constructing the forest structure of model \eqref{filt-group}. The relationship $\mathcal{G} \subseteqq \mathcal{G}'$ means that every grouping structure in $\mathcal{G}$ is also included in $\mathcal{G}'$. To make the candidate set as inclusive as possible, we require that any two different candidate sets $\mathcal{G}=\{G_{1},\ldots,G_{\ell}\}$ and $\mathcal{G}'=\{G'_{1},\ldots,G'_{\ell'}\}$ satisfy that a) $\mathcal{G}\not\subseteqq \mathcal{G}'$ and $\mathcal{G}'\not\subseteqq \mathcal{G}$, b) ${G}_1={G}'_1= \{1, 2, \ldots, p\}$, and ${G}_\ell={G}'_{\ell'}=\{\{1\}, \{2\}, \ldots, \{p\}\}$. 
The selected candidate set should ensure sufficient diversity in grouping structures to support forest construction and achieve strong predictive performance in the resulting model.
We now present the  pipeline for selecting the candidate set. 
\begin{itemize}
\item[1.] {\bf Initial Construction}: Construct all candidate sets that satisfy conditions (a) and (b) from the grouping path $\mathcal{P}$.
\item[2.] {\bf Set Evaluation}: For each candidate set, we construct the grouped model \eqref{grouped_model} based on all grouping structures in the set and estimate the prediction error using a resampling-based method (e.g., cross-validation, bootstrap).
\item[3.] {\bf Final Selection}: For each candidate set, compute the average prediction error of the models built across all the included grouping structures, and select the set that yields the lowest average prediction error. 
\end{itemize}
This selection pipeline ensures that the chosen candidate set represents an inclusive forest hierarchy spanning both the coarsest and the finest grouping structures. Candidate sets with the lowest average prediction error across grouping structures are favored because they perform well across multiple resolutions rather than excelling only at a single grouping configuration. Consequently, the grouping structures in the selected candidate set retain strong explanatory power, thereby aligning with the covariance maximization principle underlying filt-PLS.

\subsubsection{Layer-wise Structure Refinement}
\label{s3.3}
The forest structure is constructed from the selected candidate set $\widetilde{\mathcal{G}}=\{\widetilde{G}_1,\ldots,\widetilde{G}_\ell\}$ through an iterative process. If the grouping structure selected for the current layer is $\widetilde{G}_i$ based on a predefined criterion, the process continues to the next layer by sequentially testing  $\widetilde{G}_i, \ldots, \widetilde{G}_\ell$. Particularly, we sequentially examine all the grouping structures in $\widetilde{\mathcal{G}}$ in the first layer of filtration. 
To implement layer-wise refinement, we develop a prediction-aligned complexity criterion based on generalized information regularization (GIC, see e.g., \cite{ref37}).
The format of GIC criterion is: GIC = measure of model fit $+$  tuning parameter $\times$ measure of model complexity. 
In principle, if the predictors are separated into too many groups, the resulting grouping structure may fail to capture the underlying shared structure. Therefore, we penalize complex grouping structure. 

Denote a forest structure by $\mathcal{F}$, and $F_{d}\in\widetilde{\mathcal{G}}$ as the $d$-th layer's grouping structure in $\mathcal{F}$. We develop an iterative GIC criterion, with the layer-wise GIC value defined as 
$$\mbox{GIC}(F_{d})=N^{-1}\sum\limits_{n=1}^{N}(y^{[d+1]}_{n})^2+\tau_d |F_{d}|,$$ 
where $|F_{d}|$ denotes the number of groups in $F_d$, and $y^{[d+1]}_{n}$'s are the ($d+1$)-th layer's response residuals in the filt-PLS  pipeline based on the grouping structure $F_{d}$. We select $F_{d}$ to minimize $\mbox{GIC}(F_{d})$. Here, $\{\tau_d\colon d\ge1\}$ is a non-increasing sequence with respect to $d$, thus the number of groups tends to increase as $d$ increases. The number of tuning parameters $\{\tau_d\colon d\ge1\}$ increases with $d$, making tuning parameter selection increasingly complicated when the total number of layers is large. To solve this issue, we propose adopting a parametric form for $\tau_d$ (e.g., $\tau_\theta(d) = \rho \gamma^{-d+1}$, where $\theta = (\rho, \gamma)$). The parametric form significantly reduces the computational burden, as it requires selecting only $\theta$ rather than individually determining each $\tau_d$. In practice, $\theta$ can be selected by minimizing the out-of-sample prediction error of the constructed multiscale model, or by using an AIC/BIC-type criterion for the multiscale model based on the truncated shared components.
Denote $\widetilde{\mathcal{F}}$ as the identified forest structure, then we have the relationship $\widetilde{\mathcal{F}}\subseteqq \widetilde{\mathcal{G}}\subseteqq\mathcal{P}$.

We next define the stopping criterion for hierarchical refinement. In principle, the filtration should continue until the extracted layer-wise shared components no longer significantly improve the model's explanatory power. Thus, we propose the following ending criterion: we stop filtration at the $d$-th layer if $\sum_n\{(y_n^{[d]})^2-(y_n^{[d+1]})^2\}/\sum_n(y_n^{[d]})^2<e$, where $e>0$ is a small value. In addition, if $\sum_n \|X^{[d+1]}_{jn}\|^2$ becomes sufficiently small, the $j$-th predictor is removed from filtration starting at the $(d+1)$-th layer to reduce computational burden. 
The selection procedure is summarized in Algorithm \ref{alg:filtpls}.
In Step 12, we terminate the procedure when the newly extracted components explain little additional response variation over several consecutive layers. Although not essential, this step is useful in practice for preventing the selection of weak components that may appear favorable during validation solely because the corresponding multiscale model happens to yield a small out-of-sample prediction error.

\begin{algorithm}[ht]
 \caption{Iterative structure selection under GIC criterion}
  \begin{algorithmic}[1]
   \State {Initialize $y_n^{(1)}=y$ and $X_{jn}^{(1)}(t)=X_{jn}(t)$ for $j=1,\ldots,p$.}
   \For{$d=1,\ldots,D_{\max}$}
   \State {Define the active predictor set $$\mathcal{A}_d=\{j:\|X_j^{(d)}\|_F^2/\|X_j^{(1)}\|_F^2>e\}.$$}
   \If{$\mathcal{A}_d=\emptyset$}   \textbf{break.}
   \EndIf
   \State {Restrict each candidate grouping in $\mathcal{G}$ to the active predictors in $\mathcal{A}_d$.}
   \For{each candidate grouping $G$}
   \State {Fit one-layer filt-PLS using $\bigl(y_n^{(d)},\{X_{jn}^{(d)}(t)\colon j\in\mathcal{A}_d\},G\bigr)$, and obtain the residual response $y_n^{(d+1)}$ and residual predictors $\{X_{jn}^{(d+1)}\}_{j=1}^p$.}
   \State {Compute the information criterion
   \[
   \mathrm{GIC}(G)=N^{-1}\sum_{n=1}^N(y_n^{(d+1)})^2+\tau_d|G_d|,
   \]
   }
   \EndFor
   \State {Select $\widetilde F_d=\arg\min_{G\in\widetilde{\mathcal G}} \mathrm{GIC}(G)$.}
   \If{the recent relative improvements of the residual response are all below a preset threshold} \textbf{break.}
   \EndIf
   \EndFor
   \State \Return {$\widetilde{\mathcal F}=\{\widetilde F_d\}_{d\ge1}$.}
  \end{algorithmic}
  \label{alg:filtpls}
\end{algorithm}


\section{Simulation Studies}
\label{s4}
\subsection{Synthetic Setup}
In this section, we evaluate the proposed framework on synthetic data. 
The synthetic setup is designed to evaluate whether the proposed framework can recover coarse-to-fine shared structures. Exact recovery of the prespecified setup hierarchy is not the objective; rather, we aim to identify a suitable multiscale shared structure that is well aligned with the underlying setup. More importantly, our simulation results show that the identified forest structure outperforms the pre-specified setup structure, because the associated model captures the response–predictor relationship more effectively and yields better predictive performance.

\begin{figure*}[!t]
\center
\includegraphics[width=15cm]{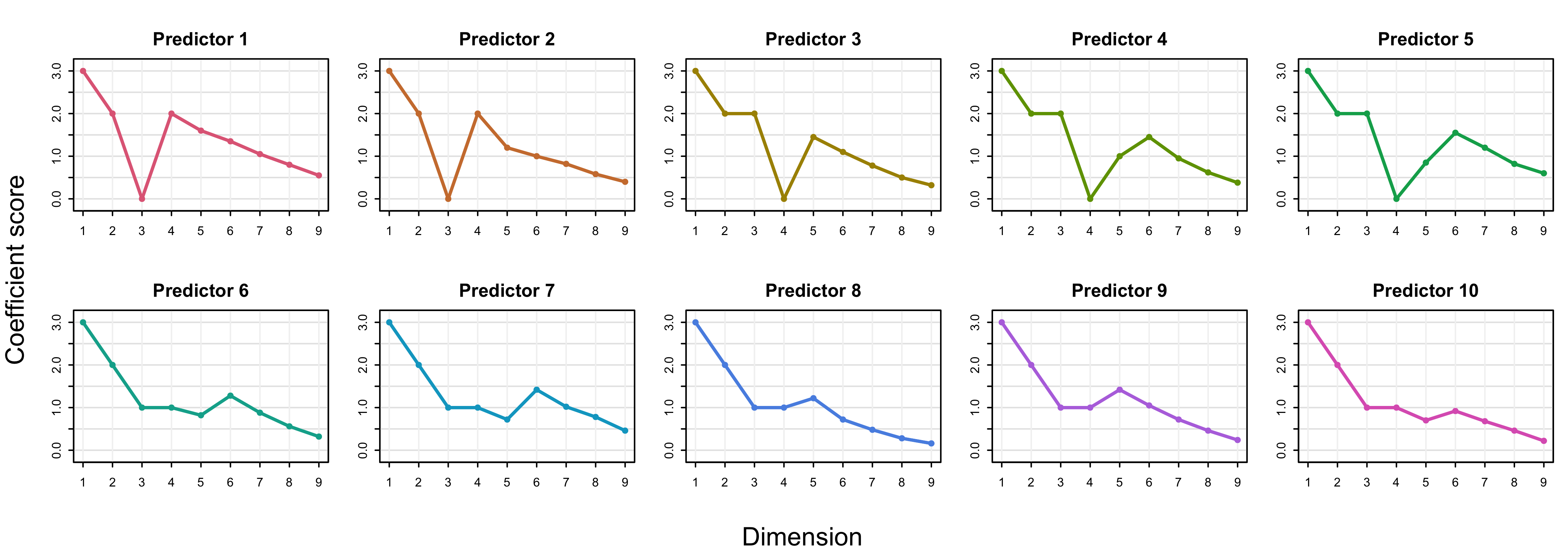}
\caption{Coefficient score profiles of the 10 functional coefficients in the synthetic setup. The first two dimensions are globally shared across all predictors, the next two encode partially shared structure, and the remaining dimensions encode predictor-specific structures, thereby inducing a coarse-to-fine multiscale organization.}
\label{coef}
\end{figure*}

We simulate $N$ samples for the following multiple functional regression model with 10 predictors $\{X_{jn}(t)\colon j=1,\ldots,10\}$,
$y_n=\sum_{j=1}^{10}\langle X_{jn},\beta_j\rangle+\epsilon_n,$ 
where $\epsilon_n\overset{i.i,d.}\sim\mathcal{N}(0,\sigma^2)$. 
The predictor and coefficient functions are simulated by the following basis expansion, 
$$X_{jn}(t)=\sum_{d=1}^D\xi_{jn,d}B_{d}(t),\ \beta_j(t)=\sum_{d=1}^Db_{jd}B_{d}(t),$$ where $\{B_{d}(t)\colon d\ge1\}$ are the Fourier basis functions. The scores $\xi_{jn,d}\stackrel{\mathrm{i.i.d.}}{\sim}\mathcal{N}(0,1.1^{-d})$ and are independent across $n$, $j$ and $d$. We impose decay on the predictor score variance to simulate the diminishing contribution of higher-order functional components.
Here we set $D=9$, and the coefficient scores $\{b_{jd}\colon j=1,\ldots,10,\ d=1,\ldots,D\}$ are displayed in Figure \ref{coef} and Table \ref{tab:beta-score}. 
\begin{table}[ht]
\centering
\caption{Coefficient scores $\{b_{jd}: j=1,\ldots,10,\ d=1,\ldots,9\}$.}
\label{tab:beta-score}
\setlength{\tabcolsep}{3.5pt}
\begin{tabular}{c| ccccccccc}
\hline
$j$ & $d=1$ & $d=2$ & $d=3$ & $d=4$ & $d=5$ & $d=6$ & $d=7$ & $d=8$ & $d=9$ \\
\hline
1  & 3.0 & 2.0 & 0.0 & 2.0 & 1.60 & 1.35 & 1.05 & 0.80 & 0.55 \\
2  & 3.0 & 2.0 & 0.0 & 2.0 & 1.20 & 1.00 & 0.82 & 0.58 & 0.40 \\
3  & 3.0 & 2.0 & 2.0 & 0.0 & 1.45 & 1.10 & 0.78 & 0.50 & 0.32 \\
4  & 3.0 & 2.0 & 2.0 & 0.0 & 1.00 & 1.45 & 0.95 & 0.62 & 0.38 \\
5  & 3.0 & 2.0 & 2.0 & 0.0 & 0.85 & 1.55 & 1.20 & 0.82 & 0.60 \\
6  & 3.0 & 2.0 & 1.0 & 1.0 & 0.82 & 1.28 & 0.88 & 0.56 & 0.32 \\
7  & 3.0 & 2.0 & 1.0 & 1.0 & 0.72 & 1.42 & 1.02 & 0.78 & 0.46 \\
8  & 3.0 & 2.0 & 1.0 & 1.0 & 1.22 & 0.72 & 0.48 & 0.28 & 0.16 \\
9  & 3.0 & 2.0 & 1.0 & 1.0 & 1.42 & 1.05 & 0.72 & 0.46 & 0.24 \\
10 & 3.0 & 2.0 & 1.0 & 1.0 & 0.70 & 0.92 & 0.68 & 0.46 & 0.22 \\
\hline
\end{tabular}
\end{table}

As illustrated in Figure \ref{coef}, the synthetic coefficient profiles are organized to exhibit a clear coarse-to-fine multiscale pattern. The first two dimensions are globally shared across all predictors, the next two induce subgroup-level sharing, and the remaining dimensions introduce predictor-specific variations. This design makes a fixed grouping configuration inadequate, while providing a natural testbed for evaluating whether the proposed framework can recover multiscale structure.
The setup hierarchy is summarized as follows:

\begin{itemize}
\item Layer 1, 2:  predictor 1--10; 

\item Layer 3, 4:  predictor 1,2; predictor 3--5; predictor 7--9; 

\item Layer 5, 6, 7, 8, 9: predictor 1; predictor 2; predictor 3; predictor 4; predictor 5; predictor 6; predictor 7; predictor 8; predictor 9; predictor 10.

\end{itemize}
The synthetic setup is not specifically designed to favor the proposed method. When shared components are weak, they are intrinsically difficult to detect from finite samples for any data-driven method. We also investigate a scenario in which the shared structures are less pronounced, where we set $\xi_{jn,d}\stackrel{\mathrm{i.i.d.}}{\sim}\mathcal{N}(0,1.1^{d-10})$ for $d=1,\ldots,9$. For each setup, we fix $N=100,\ 200$ and $\sigma=0.5,\ 1.0$, and repeat the experiment 400 times under each setup. 



\subsection{Structure Recovery and Learned Representations}
\label{B}
\begin{figure*}[!t]
\centering
\subfloat{
    \includegraphics[width=0.48\linewidth]{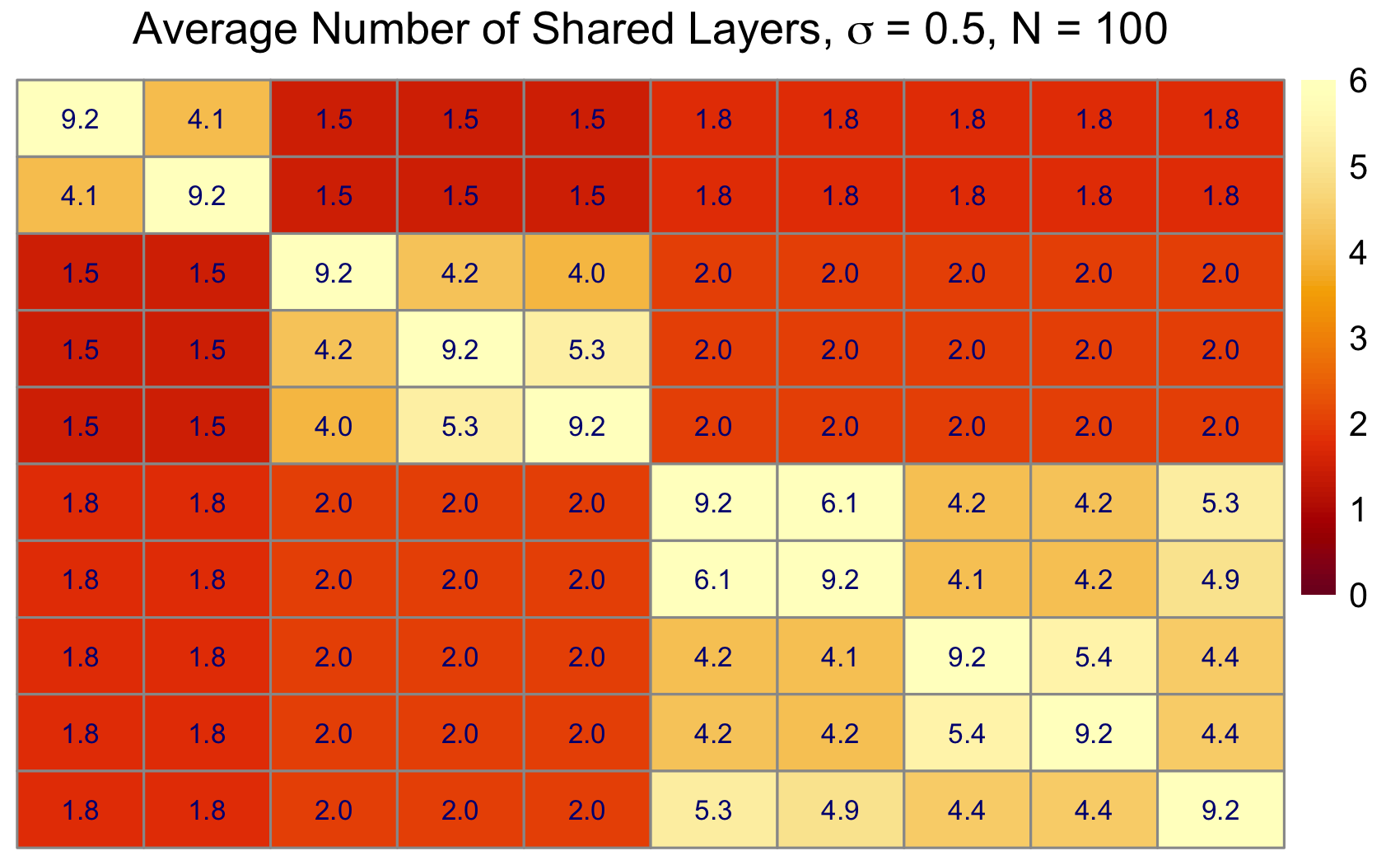}
    \label{fig:s05n100}
}
\hfill
\subfloat{
    \includegraphics[width=0.48\linewidth]{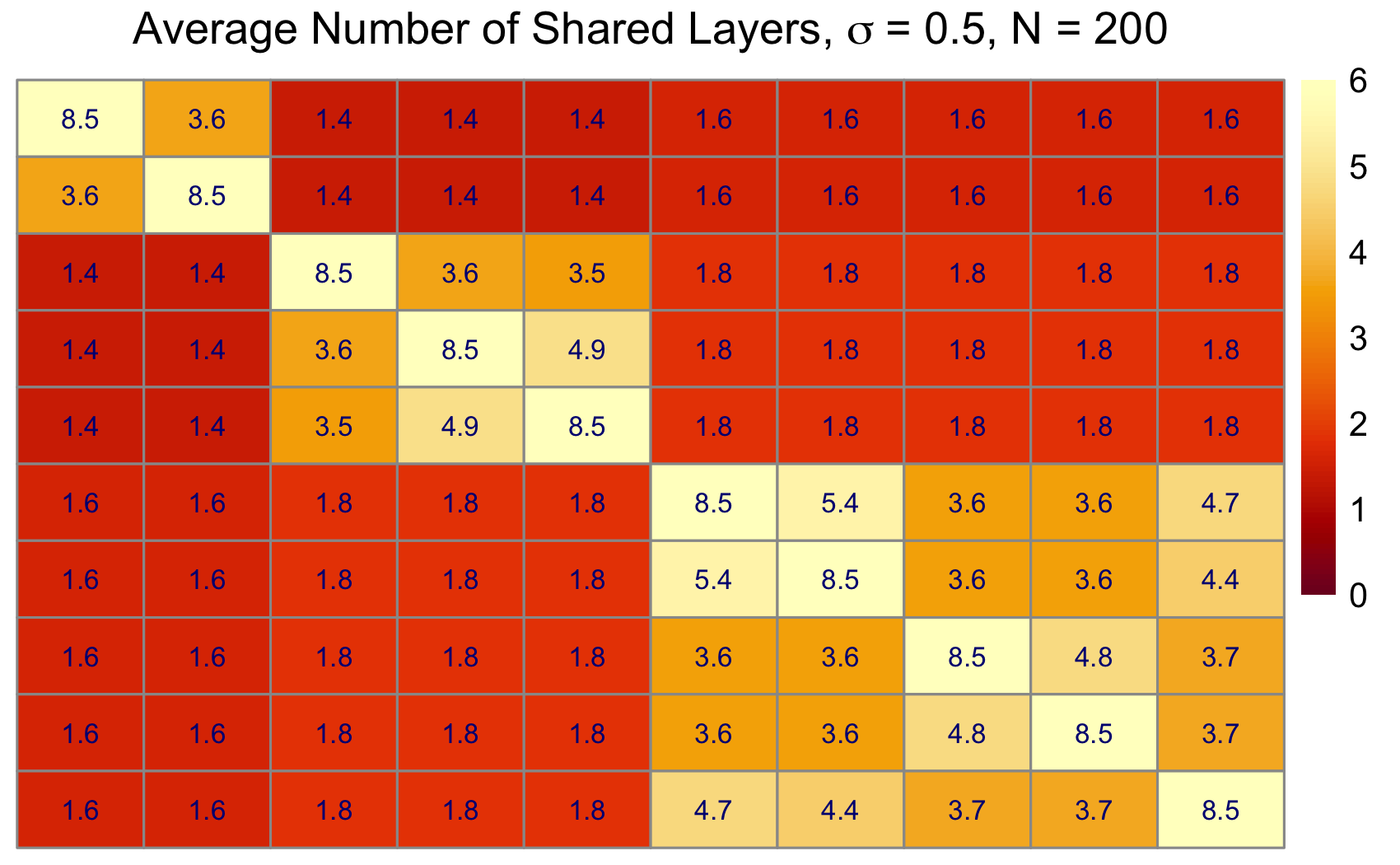}
    \label{fig:s05n300}
}

\vspace{0.2cm}

\subfloat{
    \includegraphics[width=0.48\linewidth]{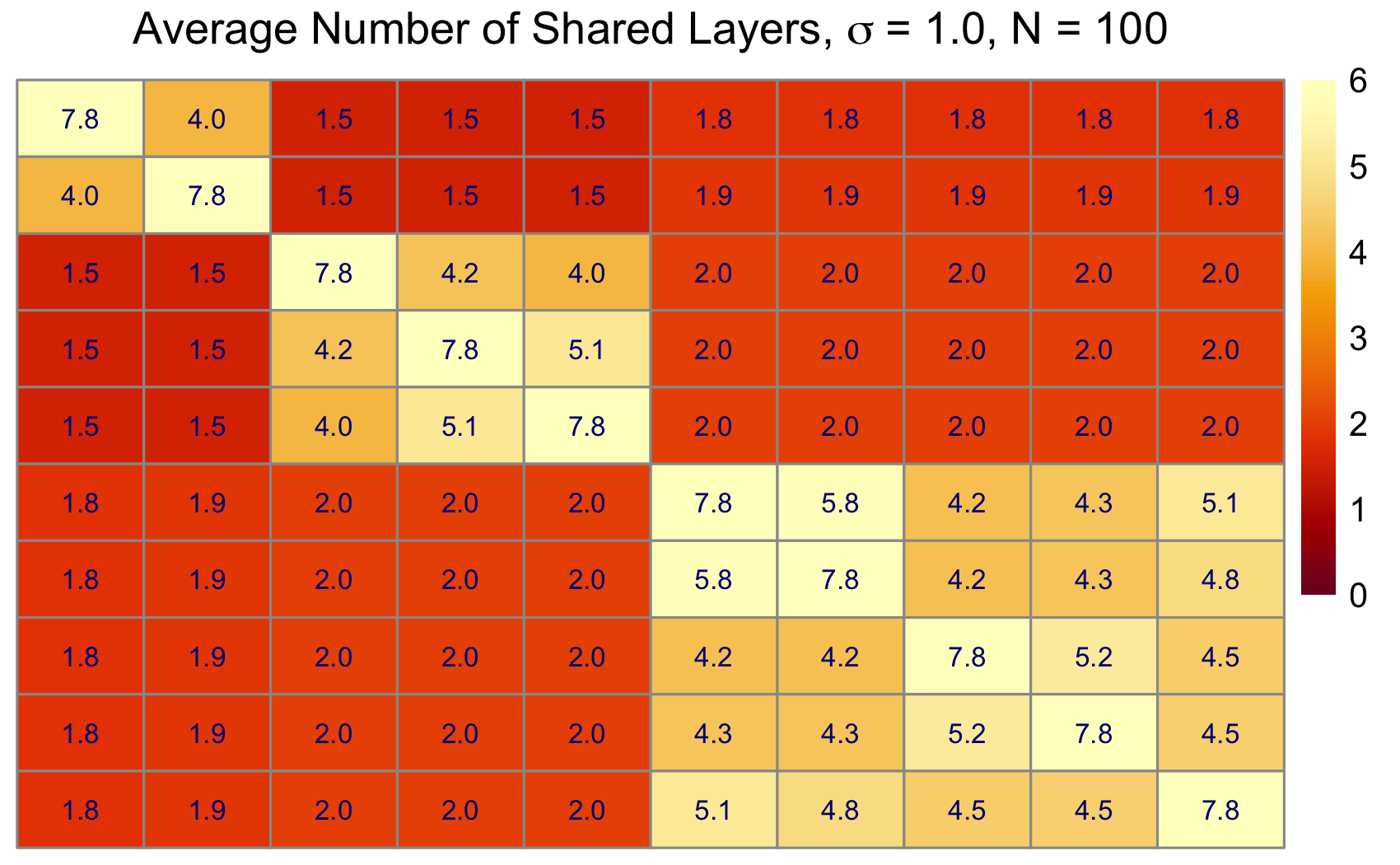}
    \label{fig:s10n100}
}
\hfill
\subfloat{
    \includegraphics[width=0.48\linewidth]{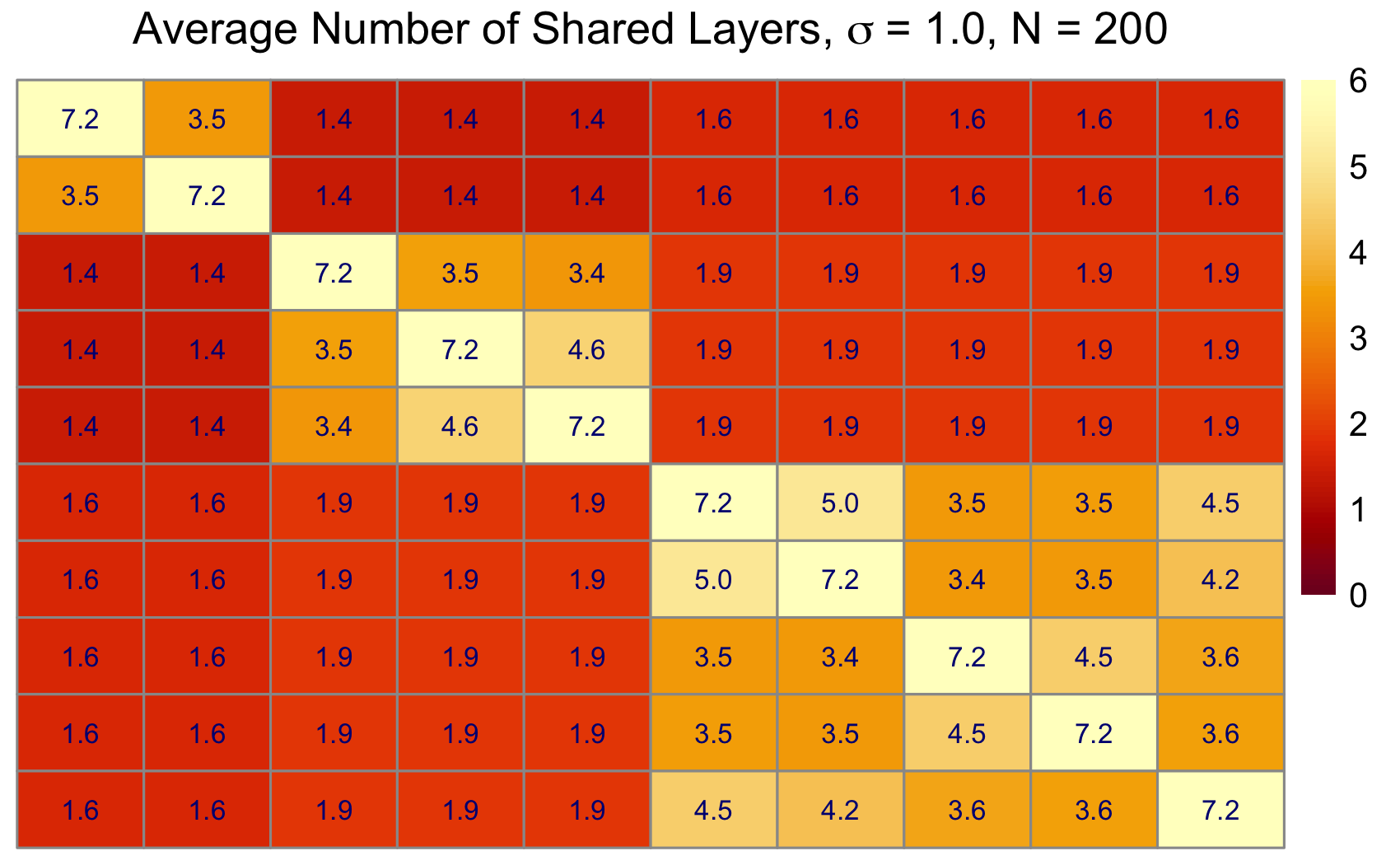}
    \label{fig:s10n300}
}

\caption{Heatmap of the average number of filtration layers in which each pair of predictors is assigned to the same group across simulation runs. The $(i,j)$ entry represents the mean number of layers for which predictors $i$ and $j$ are clustered together in the identified forest structures. Larger values indicate that the corresponding predictors tend to remain grouped together for more layers.}
\label{identified-group}
\end{figure*}

The average number of shared layers in the learned forest structures is reported in Figure \ref{identified-group}. 
Figure \ref{identified-group} reveals a clear blockwise organization in the learned hierarchies. Predictor pairs within the same intermediate-scale subgroup remain clustered together for substantially more filtration layers than pairs drawn from different subgroups, indicating that the learned representation preserves the intended coarse-to-fine shared organization. In particular, the learned hierarchy reveals global sharing in the early layers, partially shared structure among predictors 1,2, predictors 3,4,5, and predictors 6,7,8,9,10 in the intermediate layers, and progressively finer separation in the deeper layers. This pattern is consistent with the simulation design. 

\begin{figure}[!t]
\center
\subfloat{%
    \includegraphics[width=0.48\linewidth]{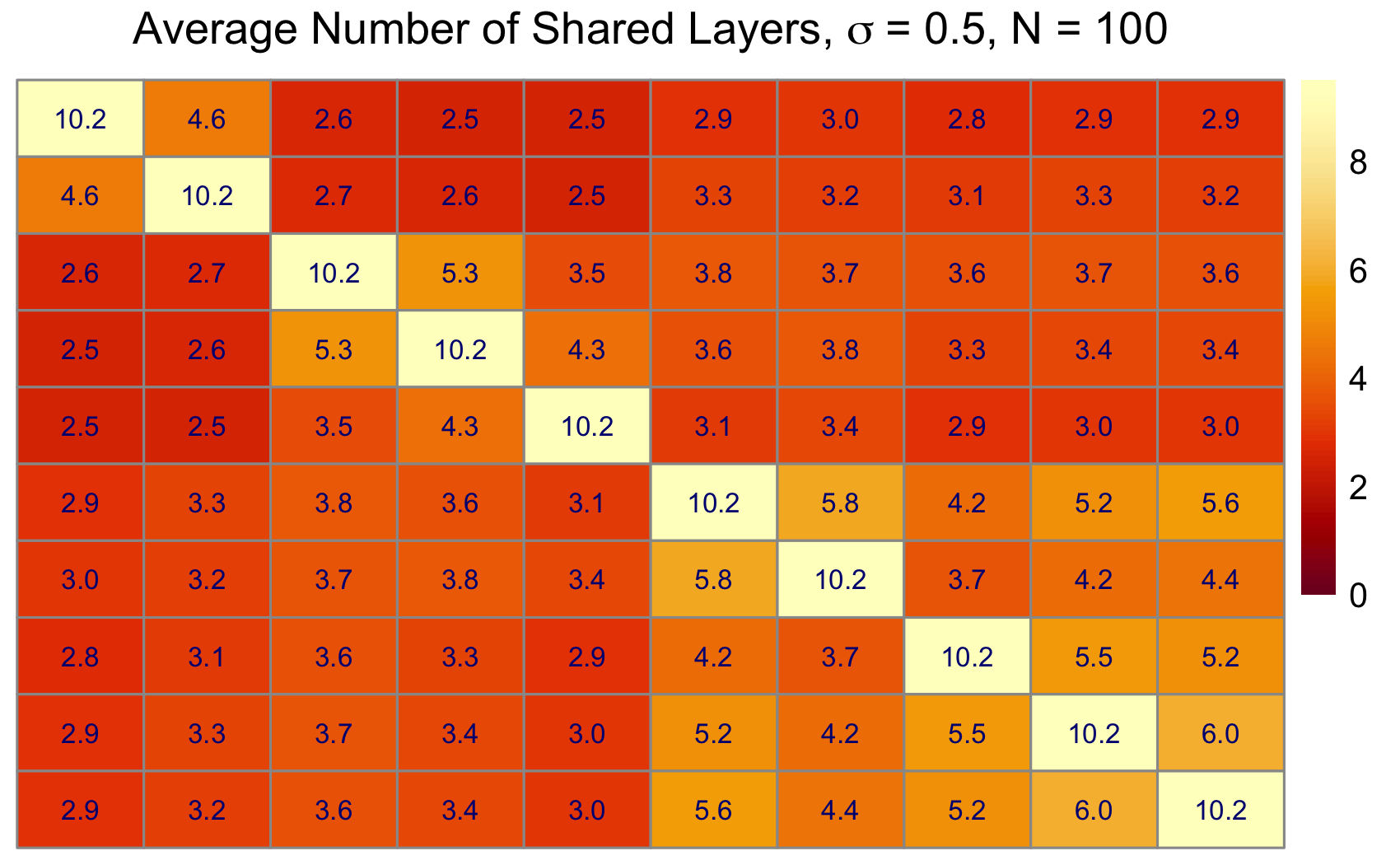}%
    \label{fig:s05n100}
}
\hfill
\subfloat{%
    \includegraphics[width=0.48\linewidth]{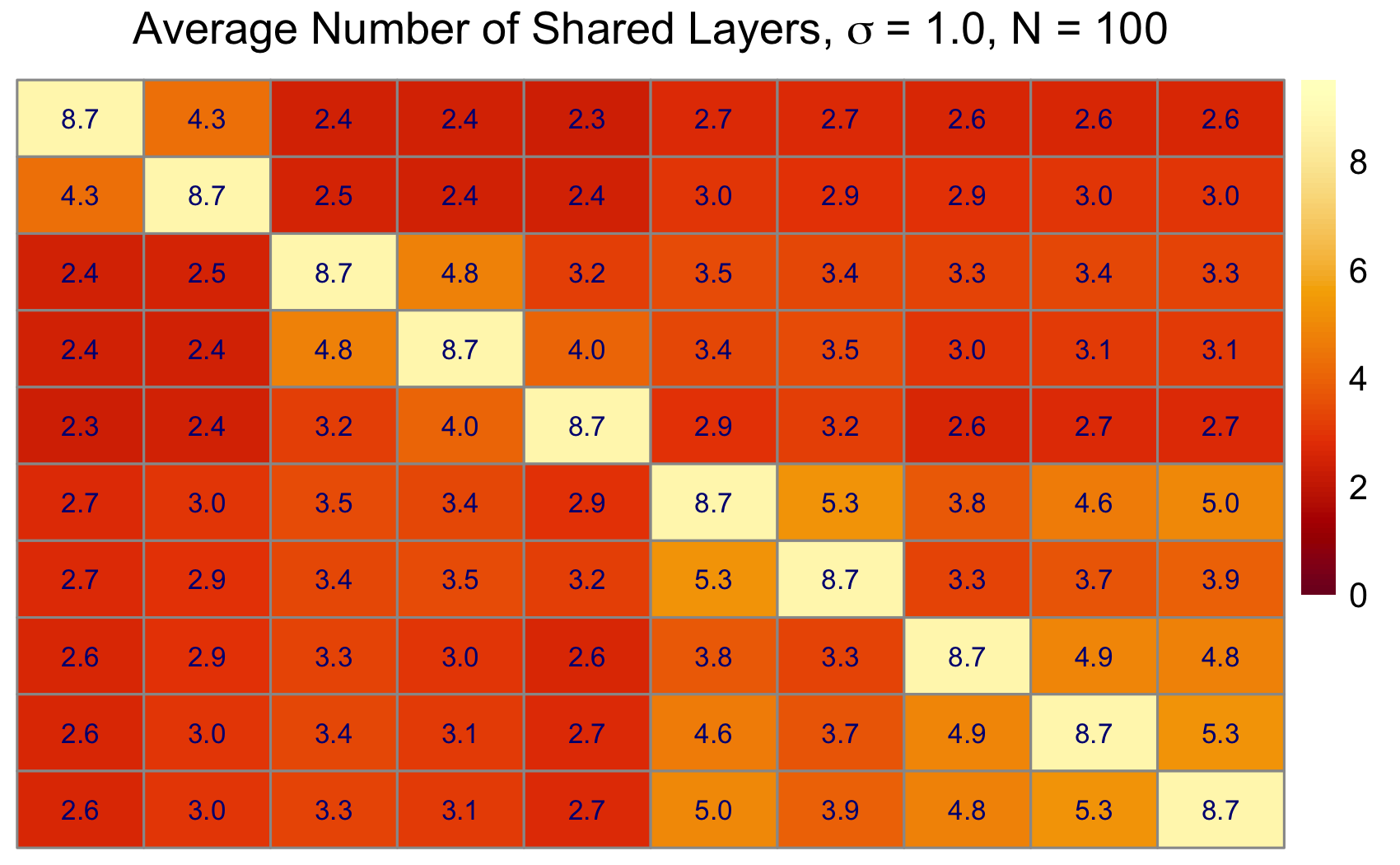}%
    \label{fig:s05n300}
}
\caption{Heatmap of the average number of filtration layers in which each pair of predictors is assigned to the same group across simulation runs under the setup where the shared components are weak.}
\label{pmse2}
\end{figure}

For the setting in which the globally and partially shared components are weak, 
the average numbers of shared layers are displayed in Figure \ref{pmse2}. The blockwise pattern is less clearly recovered than in the setting with stronger globally and partially shared signals, highlighting the difficulty of data-driven representation learning in the presence of weak shared signals. Nevertheless, the proposed multiscale model still significantly outperforms the competing methods in prediction (see Section \ref{perform}).

\subsection{Prediction Performance Comparison}
\label{perform}

\begin{figure}[!t]
\center
\includegraphics[width=12cm]{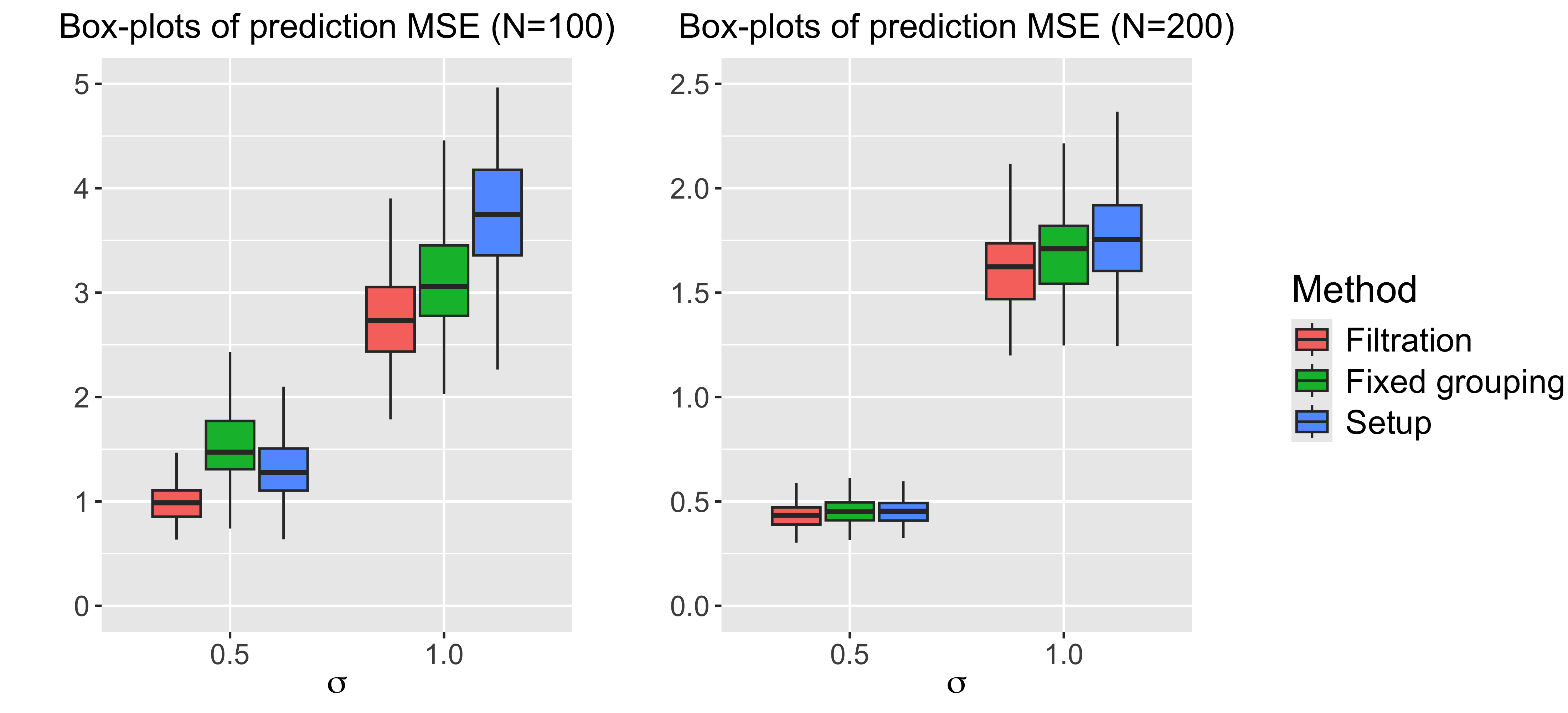}
\caption{Prediction MSEs of three competing methods in the synthetic experiments: the learned filtration hierarchy, a fixed grouping baseline, and the prespecified setup hierarchy used to generate the data.}
\label{pmse}
\end{figure}

In each simulation run, we generate either 100 or 200 samples for the training set to train the model, and 200 additional samples in the test set used for computing the prediction mean squared error (MSE). 
The boxplots of prediction MSEs are presented in Figure \ref{pmse}. It shows that the learned filtration hierarchy consistently outperforms the prespecified setup hierarchy in prediction, while also improving upon the fixed grouping baseline. 

\begin{figure}[!t]
\center
\includegraphics[width=12cm]{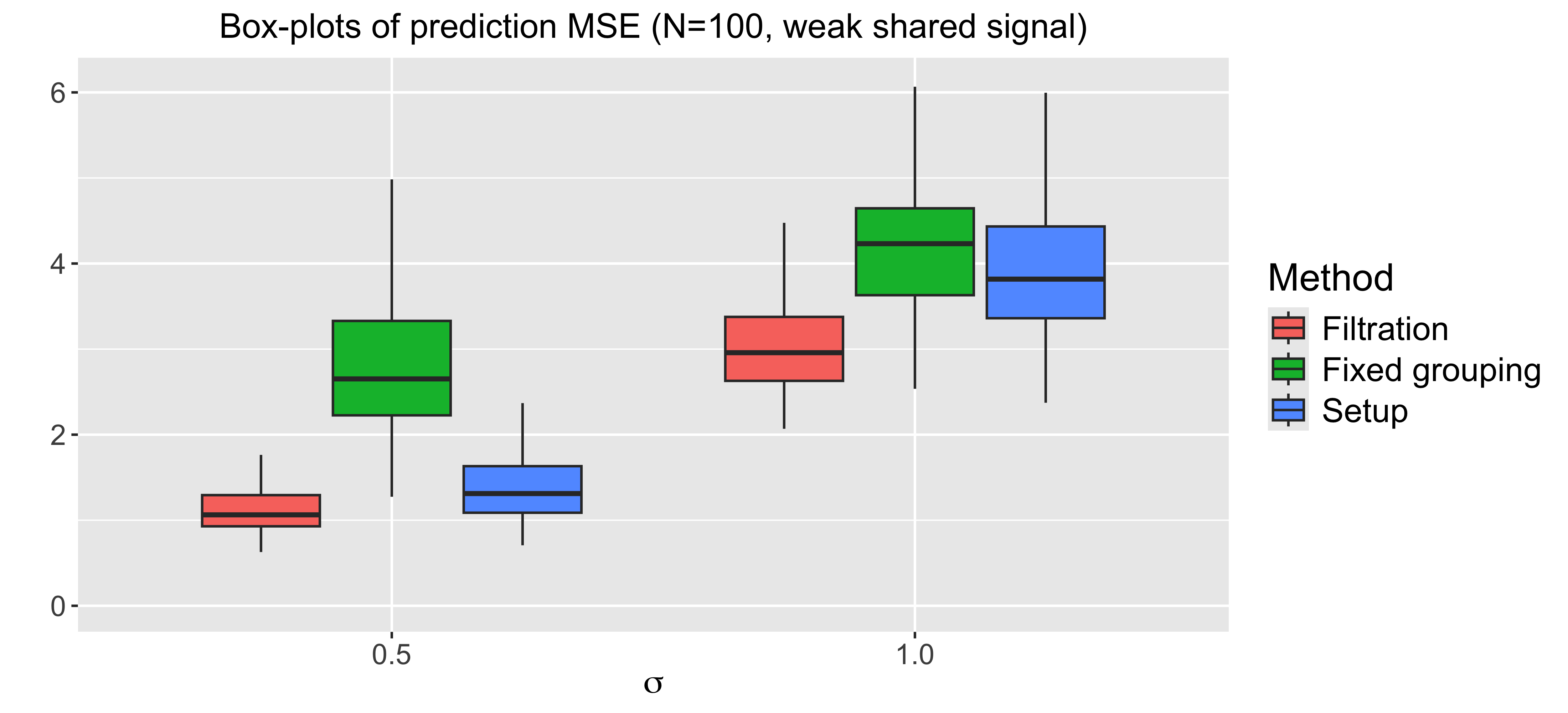}
\caption{Prediction MSEs of three competing methods in the synthetic experiments under the weak-shared-signal scenario.}
\label{pmse.r}
\end{figure}

We now explain these results. To accommodate predictor-specific effects, the fixed grouping method must use a highly refined grouping structure, which limits effective fusion, increases model complexity, and weakens prediction.
Interestingly, although the data are generated based on the pre-specified setup hierarchy, the identified shared structures yield significantly better performance. 
This improvement arises because the data-driven structures more efficiently capture the underlying mechanisms of the data, whereas the setup hierarchy does not necessarily provide the most efficient representation of the underlying multiscale organization. Thus, the learned structures lead to more parsimonious representation and reduce the estimation error. 
In addition, the filtration-based approach still consistently achieves the best and most robust predictive performance under the scenario where the shared signals are weak (see Figure \ref{pmse.r}).

Overall, the simulation results support the conclusion that hierarchical structure learning is essential when shared structures evolve across multiple resolutions, and that the learned multiscale structures capture the underlying data variation more effectively, as reflected in their superior predictive performance.

\section{Filtrated Kinematic Connectivity Analysis\\ in Lower limb Joint Effective Age Evaluation}
\label{s5}
This analysis is driven by the scientific interest in the intersection of chronic risk communication and gait analysis. An emerging concept, referred to as “effective age,” has garnered increasing attention (see e.g., \cite{ref1,ref32}). Effective age represents the average age level of healthy individuals who share the same risk profile as a person being evaluated. In many cases, a person’s health status is more accurately reflected by their effective age rather than their chronological age, as it accounts for the impact of chronic disease risks and overall well-being. Evaluating effective age through gait analysis is a vital component in the proactive management of joint health, facilitating early detection of degeneration, customization of rehabilitation strategies, and the promotion of sustained mobility.


We apply the proposed framework to lower-limb angular kinematics to study age-related connectivity patterns in gait using the data collected from the experiment conducted by \cite{ref13}. Here, ``connectivity'' refers to age-associated homogeneous patterns shared across multiple kinematic trajectories, that is, response-relevant shared structure rather than merely pairwise dependence between joints. Our goals are threefold: 1) to evaluate whether the proposed framework can identify multiscale gait connectivity patterns for lower-limb effective age assessment; 2) to determine which shared and predictor-specific kinematic components are informative for this evaluation task; and 3) to assess whether these multiscale shared components improve age evaluation performance compared with methods based on fixed shared structures.


The study involves 42 healthy volunteers aged between 21 and 84, all of whom had no lower-extremity injuries in the six months prior to participation and no orthopedic or neurological conditions affecting their gait. Since the number of participants is limited, it is crucial to avoid building models with high complexity. Studying the association between chronological age and gait angular kinematics in healthy individuals is essential, as it serves as a baseline for assessing new subjects. Although chronological age and effective age do not always align in healthy individuals, particularly in those engaging in regular exercise that can reduce biological age by several years, the phenomenon is not universal and should not be assumed to occur in most healthy people. Data were collected using standard gait-analysis  pipelines in a 10×12-m room at the Laboratory of Biomechanics and Motor Control at the Federal University of ABC, Brazil. This facility is equipped with a motion-capture system that had 12 cameras, five force platforms embedded in the floor, and a dual-belt instrumented treadmill. Our study focuses on treadmill walking data, as they offer greater consistency and are not influenced by terrain variations. The angular kinematics data were recorded at various gait speeds, categorized into eight levels ranging from the slowest to the fastest speeds. 

\begin{figure}[tb]
\center
\includegraphics[width=15cm]{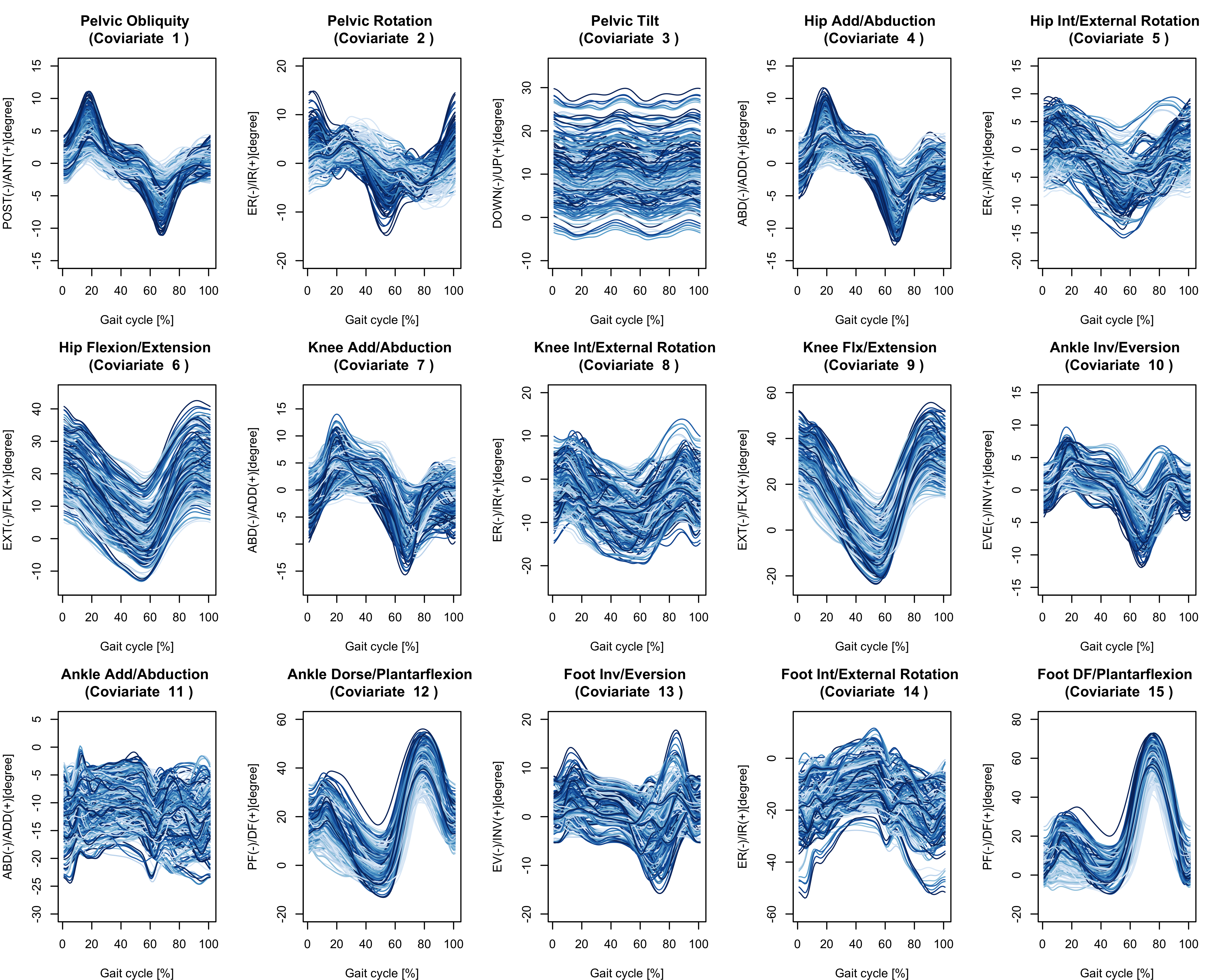}
\caption{Angular kinematic trajectories averaged across the left and right sides, with curve colors indicating walking speed (lighter color indicates slower speed).}
\label{cov}
\end{figure}

We average the angular kinematic trajectories from the left and right sides to reduce model complexity while retaining focus on kinematic connectivity. Consequently, there are 15 kinematic trajectories per gait cycle, each serving as a functional predictor, including pelvic obliquity, pelvic rotation, pelvic tilt, hip add/abduction, hip int/external rotation, hip flexion/extension, knee add/abduction, knee int/external rotation, knee flx/extension, ankle inv/eversion, ankle add/ abduction, ankle dorsi/plantarflexion, foot inv/eversion,  foot int/external rotation and foot DF/plantarflexion angles (see Figure \ref{cov}). To improve model interpretability, we centralize the log-transformed age and use it as the response $y_n$, and standardize the angular kinematic curves as follows, $\widetilde{X}_{jn}(t)=\{X_{jn}(t)-\overline{X}_{j}(t)\}\{N^{-1}\sum_{n=1}^N\|X_{jn}(t)-\overline{X}_{j}(t)\|^2\}^{-1/2},\ j=1,\ldots,15$, where $\overline{X}_{j}(t)=N^{-1}\sum_{n=1}^NX_{jn}(t)$. 
This preprocessing ensures that the learned grouping structure reflects shared temporal and structural movement patterns rather than differences in raw angular scale.

Here, we focus on the comfortable walking speed, specifically within the range of 100–115\% of normal walking speed. 
We employ a Monte Carlo cross-validation approach (see \cite{ref34}) to select the tuning parameters in the GIC criterion. Specifically, the samples are randomly split into a training set (80\%) and a test set (20\%). The model trained on the training set is used to predict the responses in the test set. We repeat this  pipeline 1000 times to obtain the average prediction MSE. The tuning parameters $\theta$ associated with the minimal prediction MSE is then used for forest structure identification, resulting in the forest structure displayed in Figure \ref{commu}. The shared structures across different filtration layers are shown below:
\begin{itemize}
\item[$\widetilde{F}_1$.] $\mathcal{K}_{1,1}$: predictor 1, 4;  $\mathcal{K}_{1,2}$: predictor $2, 3, 5,\ldots,15$; 

\item[$\widetilde{F}_2$.] $\mathcal{K}_{2,1}$: predictor 1; $\mathcal{K}_{2,2}$: predictor $2, 3, 5, 6, 8, \ldots, 15$; $\mathcal{K}_{2,3}$: predictor 4; $\mathcal{K}_{2,4}$: predictor 7; 

\item[$\widetilde{F}_3$.] $\mathcal{K}_{3,i}$: predictor $i$, for $i=1,\ldots,p$;

\item[$\widetilde{F}_4$.] $\mathcal{K}_{4,1}$: predictor 2; $\mathcal{K}_{4,2}$: predictor 5; $\mathcal{K}_{4,3}$: 
predictor 10; $\mathcal{K}_{4,4}$: predictor 11; $\mathcal{K}_{4,5}$: predictor 13; $\mathcal{K}_{4,6}$: predictor 14; $\mathcal{K}_{4,7}$: predictor 15;

\item[$\widetilde{F}_5$.] $\mathcal{K}_{5,1}$: predictor 11.
\end{itemize}


\begin{figure}[tb]
\center
\includegraphics[width=15cm]{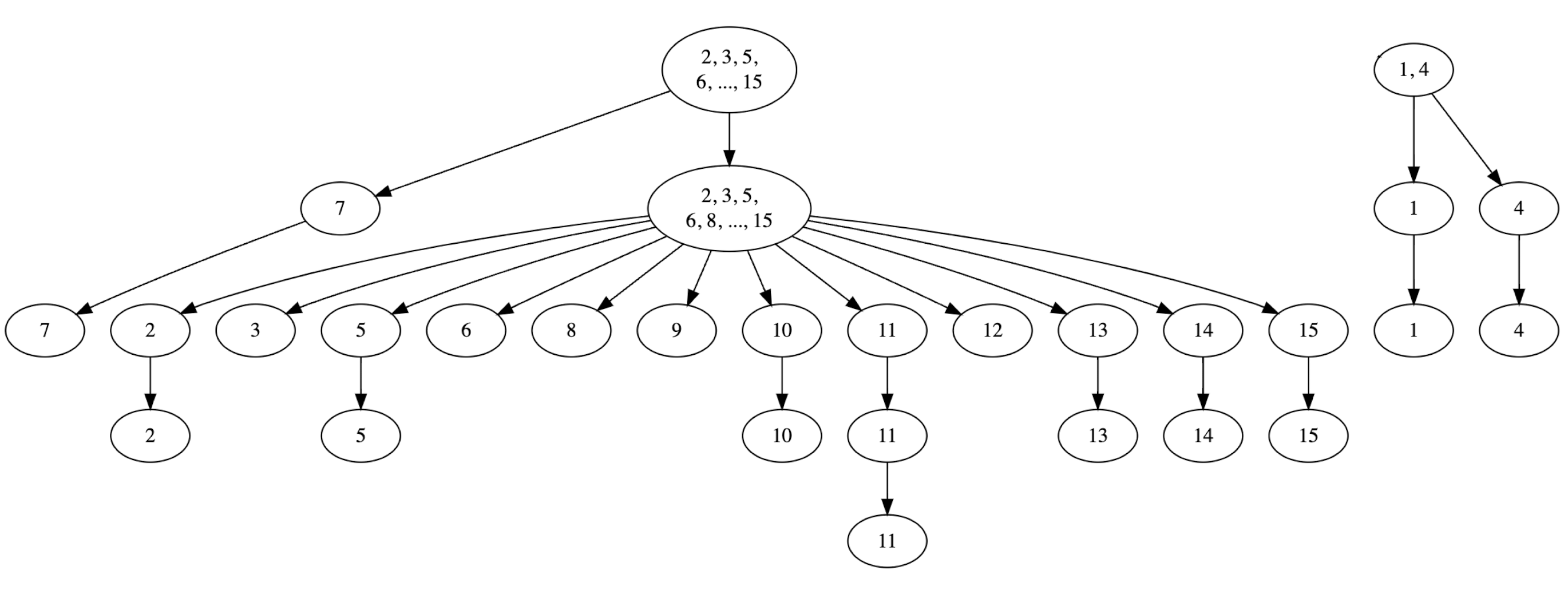}
\caption{The identified forest structure under 100–115\% normal walking speed (the numbers in the circles represent the predictor indices). There are five layers in total. Most predictors are removed from Layers 4 and 5, as the extracted shared components already account for a substantial proportion of their variation.
}
\label{commu}
\end{figure}
\begin{figure}[!t]
\centering
\includegraphics[width=5in]{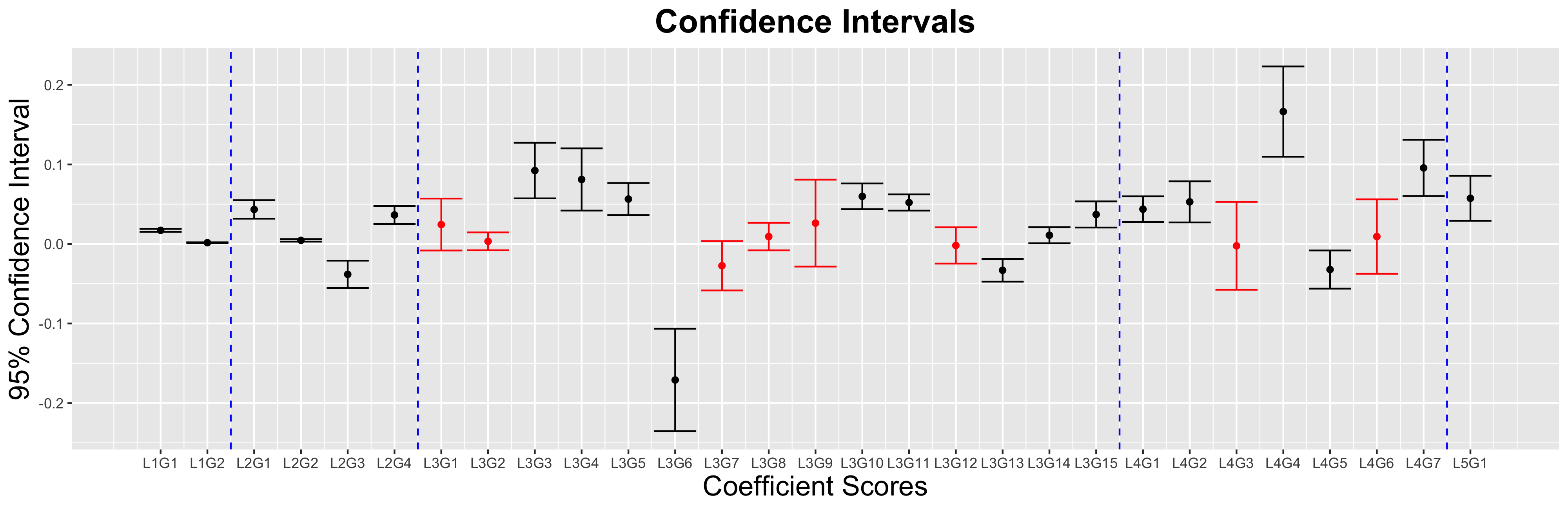}\\%
\includegraphics[width=5in]{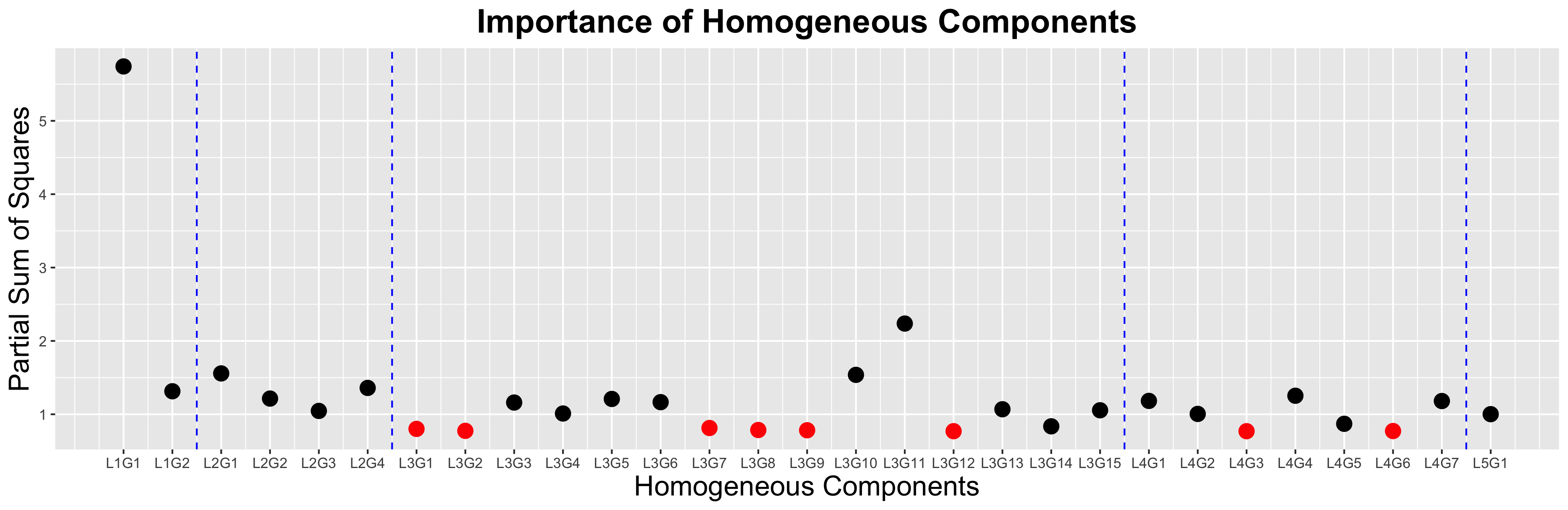}%
\caption{The confidence intervals for coefficient scores and the PSS values. Different layers are separated by blue dashed lines. L$_d$G$_i$ represents the $i$-th group in the $d$-th layer, same as $\mathcal{K}_{d,i}$. The insignificant shared components are marked in red.}
\label{coefsc}
\end{figure}

 
Define the partial sum of squares (PSS) of the layer-wise shared components of layer $d$ group $i$ as $\mbox{RSS}_{d,i}$, where  $\mbox{RSS}_{d,i}$ represents the residual sum of squares of the reduced model with the corresponding shared component  omitted from the full model (all the identified shared components are involved). We use the partial sum of squares to evaluate the importance of the shared components, and a larger value of PSS indicates greater importance.
We display the 95\% confidence interval of the coefficient scores $\{a_{d,i}\colon i,d\ge1\}$ and the partial sum of squares of all the identified shared components in Figure \ref{coefsc}. 
{\it Notably, all coefficient scores in the first two layers are significantly nonzero, confirming our hypothesis that kinematic connectivity imparts a consistent age-associated pattern across different kinematic trajectories.}

\begin{figure}[tb]
\center
\includegraphics[width=15cm]{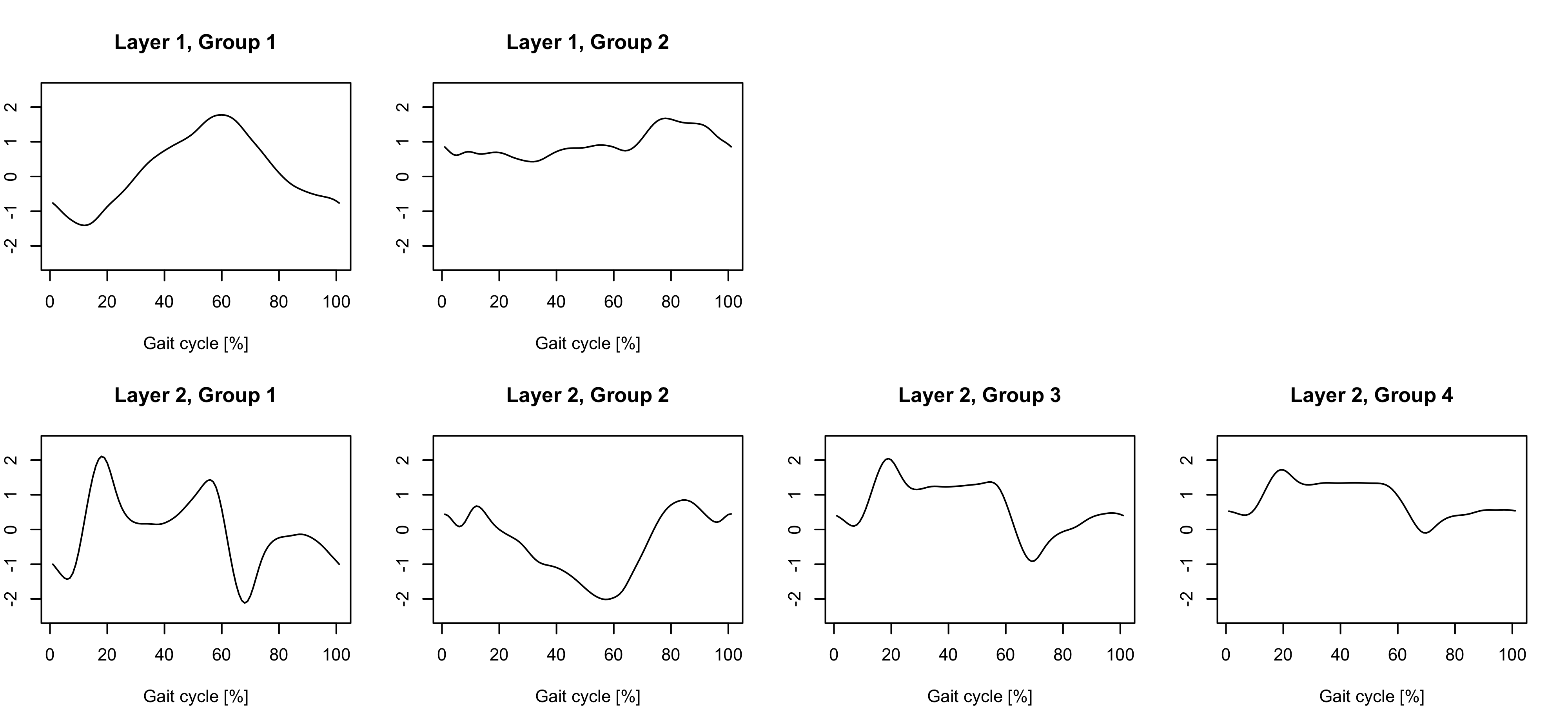}
\caption{The partial least square basis functions in the first two layers.}
\label{plsbasis}
\end{figure}

The dominant shared structure in the age--kinematics association is captured by the shared components in $\mathcal{K}_{1,1}$, $\mathcal{K}_{1,2}$, and $\mathcal{K}_{2,2}$. 
In the first hierarchical layer, predictor 1 and 4, representing pelvic obliquity and hip adduction/abduction in the frontal plane, are separated from the other angular kinematic trajectories. 
The filt-fPLS basis functions of the two groups in the first layer (see Figure \ref{plsbasis}) differ markedly, underscoring that pelvic and hip frontal-plane dynamics are associated with age in distinct ways.
This reflects the corresponding collective neuromuscular and postural adaptations: during aging, decreased mediolateral balance control and altered weight distribution drive compensatory adjustments in pelvis and hip frontal-plane motion. 
The corresponding PSS value is considerably higher than those of other layer-wise shared components, indicating that pelvic and hip frontal-plane motion coordinate as the primary kinematic feature for assessing joint effective age.
Meanwhile, we find that the predictor-specific kinematic features of knee internal/external rotation, knee flexion/extension, and ankle dorsiflexion/plantarflexion are non-significant in their associations with age. Their associations with age are adequately captured in the first two filtration layers through the components shared with other kinematic trajectories, and their predictor-specific components do not distinguish gait characteristics between younger and older individuals. Interestingly, the PSS value of the shared components in $\mathcal{K}_{3,11}$, corresponding to the predictor-specific feature of ankle adduction/abduction dynamics exhibits a pronounced spike. Additionally, the three layer-wise shared components in $\mathcal{K}_{3,11}$, $\mathcal{K}_{4,4}$, and $\mathcal{K}_{5,1}$, which also correspond to the predictor-specific features of ankle adduction/abduction, all show significant effects. These findings suggest that ankle adduction/abduction kinematics also serve as a particularly informative biomarker for evaluating effective age, and its individual adaptation to aging is non-negligible.


The filt-PLS basis functions $\{\psi_{d,i}(t)\colon d \ge 1,\ i = 1,\ldots,m_d\}$ are displayed in Figure \ref{plsbasis}, and provide interpretable insights into how the shared components relate to age over the gait cycle. Figures \ref{coefsc} and \ref{plsbasis} provide a clear interpretation of how the corresponding kinematic trajectories are associated with age. For example, in the first layer, the shared component in Group 1 is primarily associated with pelvis obliquity and hip adduction/abduction. It exhibits a positive relationship with these angular kinematic trajectories during 40–70\% of the gait cycle and a negative relationship during the remaining phases, forming a bell-shaped pattern. Combining the result that the associated coefficient score is significantly positive, we conclude that the 40–70\% period of the aggregated kinematic trajectories is positively associated with age, whereas the remaining phases exhibit a negative association. In contrast, the shared components in Group 2 are positively driven by all other angular kinematic trajectories throughout the entire gait cycle and exhibit positive association with the response. The corresponding filt-PLS basis is approximately a constant function, capturing the mean level of the aggregated kinematic trajectories.

\begin{figure}[tb]
\center
\includegraphics[width=15cm]{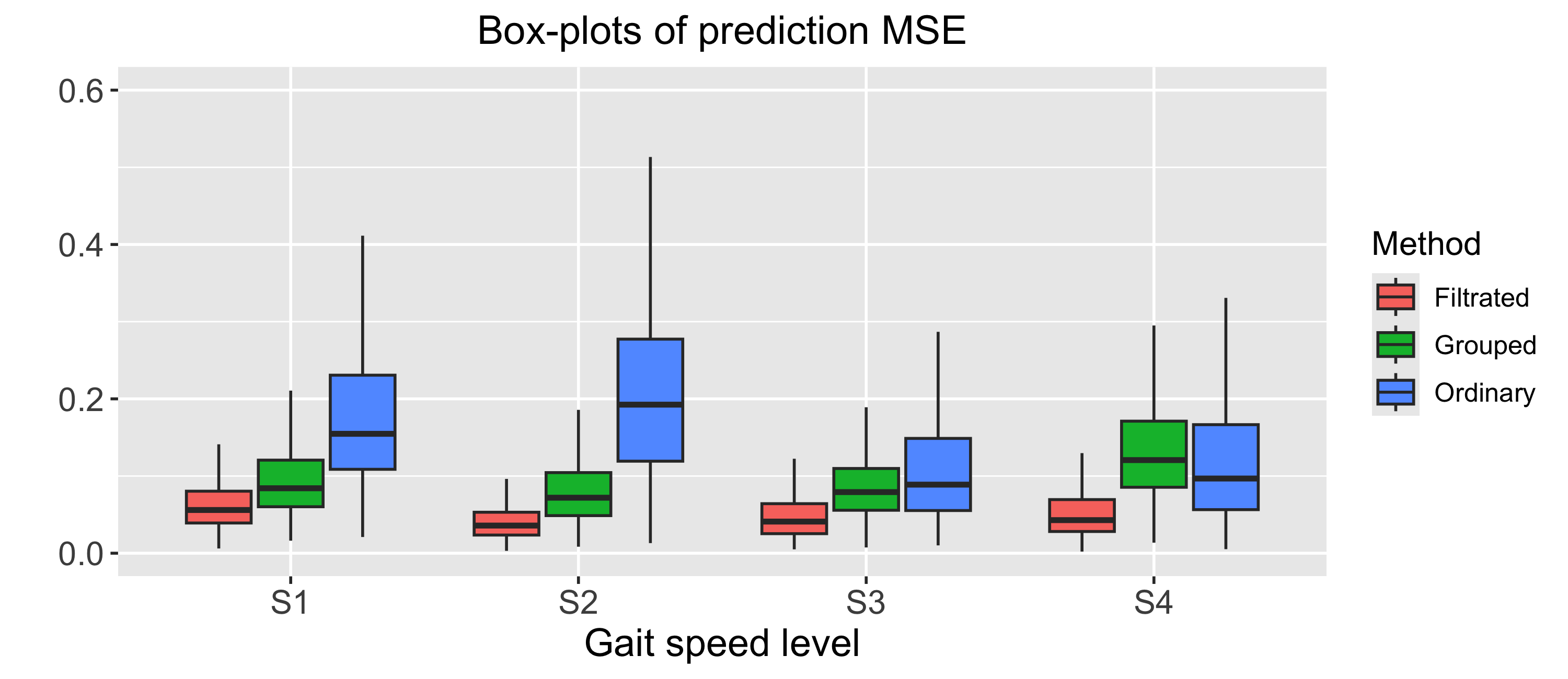}
\caption{Box-plots of MSEs of different methods: the proposed multiscale learning method (Filtrated), the ordinary functional regression model (Ordinary) and the grouped functional regression model (Grouped). The variance of the log-transformed age under the four gait speed levels is 0.19, 0.19, 0.19, and 0.17, respectively.}
\label{pmse-rd}
\end{figure}

We benchmark the performance of the new method against two established alternatives: the ordinary functional regression model \eqref{mlr} and the grouped functional regression model \eqref{grouped_model}, through comparing the prediction error for age. Prediction accuracy is evaluated using Monte Carlo cross-validation. The participants were required to walk on treadmill under different gait speeds, and we categorize them into four groups: 40–55\% (S1), 70–85\% (S2), 100–115\% (S3), and 130–145\% (S4) of the normal walking speed. Then we establish multiple functional linear models under different gait speed groups with response $y^{(s)}_n$ and predictors $\widetilde{X}^{(s)}_{jn}(t)$, where the superscript $(s)$ indicates the gait speed levels, $y^{(s)}_n=\sum_{j\ge1}\langle\widetilde{X}^{(s)}_{jn},\beta^{(s)}_j\rangle+\epsilon_n^{(s)}.$ We repeat the Monte-Carlo sampling 1000 times, and the resulting mean squared errors (MSEs) are displayed in Figure \ref{pmse-rd}. The multiscale model achieves optimal predictive performance at 70--85\% of normal walking speed, indicating that assessments conducted at a slightly reduced pace relative to an individual’s comfortable walking speed yield more precise effective age assessment.

Across all gait speed levels, the multiscale model consistently delivers the lowest and most robust MSEs.
The ordinary model \eqref{mlr} underperforms due to its high parameter complexity, which leads to increased assessment uncertainty.
The grouped model \eqref{grouped_model} demonstrates improvements over the ordinary model; however, it remains inferior to the proposed multiscale model. This shortfall stems from its failure to accommodate heterogeneous, predictor-specific associations between individual predictors and response. By grouping predictors consistently over all dimensions, the grouped model masks their distinct contributions.
In contrast, the multiscale model’s hierarchical grouping structure effectively distinguishes shared versus predictor-specific effects, allowing for much more nuanced modeling of age-kinematics associations and significantly enhancing both interpretability and evaluation performance.

\section{Conclusion and Discussion}
\label{s6}
We introduced a filtration-based framework for learning multiscale shared structures from multiple functional predictors.
The proposed approach captures globally shared, partially shared, and predictor-specific effects through a hierarchical multiscale representation, enabling structured learning from multiple functional predictors with improved interpretability and predictive performance. 
The advantage of the filtration framework is not confined to exact recovery at any single layer. Instead, it lies in the hierarchical organization learned across multiple layers, through which the method can capture the dominant multiscale shared pattern. From this perspective, the key point is not whether one particular layer matches the nominal oracle grouping, but whether the learned hierarchy as a whole faithfully reflects the underlying coarse-to-fine shared structure and supports accurate and interpretable representation learning. 

Beyond its empirical performance, the proposed framework introduces filtration as a general principle for hierarchical structure learning in multivariate functional data. The resulting representation is data-adaptive and multiscale, allowing predictor interactions to be organized from coarse shared patterns to fine predictor-specific effects within a unified framework. This perspective may be useful more broadly for structured learning problems in which dependency patterns evolve across scales.

To our knowledge, this is the first work to study multiple functional regression from the perspective of multiscale shared-structure learning under a filtration-based hierarchy.
This perspective opens several directions for future research. For example, filtration ideas may be extended to other learning problems involving multiple multidimensional objects, including factor models, structured dimension reduction, and nonlinear regression with multiple functional inputs. It is also of interest to study how filtrated structures vary across populations, tasks, or operating conditions, and how such differences can be exploited for comparative learning and representation analysis. Related ideas have appeared in network analysis for stable multiscale comparison across filtration levels \cite{ref19}, and our results suggest that a similar perspective can be fruitful much more broadly.


\begin{thebibliography}{99}

\bibitem{ref1}
A. S. Anderson and R. F. Loeser, ``Why is osteoarthritis an age-related disease?,'' \textit{Best Pract. Res. Clin. Rheumatol.}, vol. 24, pp. 15--26, 2010.


\bibitem{ref3}
Y. Cheng, J. Q. Shi, and J. Eyre, ``Nonlinear mixed-effects scalar-on-function models and variable selection,'' \textit{Stat. Comput.}, vol. 30, pp. 129--140, 2020.



\bibitem{ref6}
J.-M. Chiou, Y.-F. Yang, and Y.-T. Chen, ``Multivariate functional linear regression and prediction,'' \textit{J. Multivar. Anal.}, vol. 146, pp. 301--312, 2016.

\bibitem{ref7}
J. Dannenmaier, C. Kaltenbach, T. Kolle, and G. Krischak, ``Application of functional data analysis to explore movements,'' \textit{Gait Posture}, vol. 77, pp. 182--189, 2020.

\bibitem{ref8}
A. Delaigle and P. Hall, ``Methodology and theory for partial least squares applied to functional data,'' \textit{Ann. Statist.}, vol. 40, pp. 322--352, 2012.



\bibitem{ref11}
J. Fan and R. Li, ``Variable selection via nonconcave penalized likelihood,'' \textit{J. Amer. Statist. Assoc.}, vol. 96, pp. 1348--1360, 2001.




\bibitem{ref13}
C. A. Fukuchi, R. K. Fukuchi, and M. Duarte, ``A public dataset of overground and treadmill walking kinematics and kinetics in healthy individuals,'' \textit{PeerJ}, vol. 6, p. e4640, 2018.



\bibitem{ref15}
S. Jiao and N.-H. Chan, ``Coefficient shape alignment in multiple functional linear regression,'' \textit{J. Amer. Statist. Assoc.}, vol. 119, pp. 1--14, 2024.

\bibitem{ref16}
S. Jiao, R. Frostig, and H. Ombao, ``Filtrated common functional principal component analysis of multigroup functional data,'' \textit{Ann. Appl. Stat.}, vol. 18, pp. 1160--1177, 2024.



\bibitem{ref18}
Z. T. Ke, J. Fan, and Y. Wu, ``Homogeneity pursuit,'' \textit{J. Amer. Statist. Assoc.}, vol. 110, pp. 175--194, 2015.

\bibitem{ref17}
R. M. Kay, S. Dennis, S. Rethlefsen, D. L. Skaggs, and V. T. Tolo, ``Impact of postoperative gait analysis on orthopaedic care,'' \textit{Clin. Orthop. Relat. Res.}, vol. 374, pp. 259--264, 2000.

\bibitem{ref45}
E. D. Kolaczyk and H. Huang,
``Multiscale Statistical Models for Hierarchical Spatial Aggregation,''
\textit{Geographical Analysis},
vol. 33, no. 2, pp. 95--118, 2001.

\bibitem{ref19}
H. Lee, M. K. Chung, H. Kang, B.-N. Kim, and D. S. Lee,
``Computing the shape of brain networks using graph filtration and Gromov--Hausdorff metric,''
in \textit{Lecture Notes in Computer Science},
vol. 6892, pp. 302--309, 2011.

\bibitem{ref39}
S. Lei, C. Hao, and M. Qi, ``Multi-Scale Feature for Recognition,'' in \textit{Proc. Int. Conf. Electron. Comput. Technol. (ICECT)}, pp. 277--280, 2009.






\bibitem{ref46}
E. F. Lock, K. A. Hoadley, J. S. Marron, and A. B. Nobel,
``Joint and Individual Variation Explained (JIVE) for Integrated Analysis of Multiple Data Types,''
\textit{Ann. Appl. Stat.},
vol. 7, no. 1, pp. 523--542, 2013.

\bibitem{ref22}
B. Lofter{\o}d, T. Terjesen, I. Skaaret, A.-B. Huse, and R. Jahnsen, ``Preoperative gait analysis has a substantial effect on orthopedic decision making in children with cerebral palsy,'' \textit{Acta Orthop.}, vol. 78, pp. 74--80, 2007.




\bibitem{ref23}
S. Ma and J. Huang, ``A concave pairwise fusion approach to subgroup analysis,'' \textit{J. Amer. Statist. Assoc.}, vol. 112, pp. 410--423, 2017.

\bibitem{ref24}
J. S. Morris, ``Functional regression,'' \textit{Annu. Rev. Stat. Appl.}, vol. 2, pp. 321--359, 2015.


\bibitem{ref26}
H.-G. M\"uller and F. Yao, ``Functional additive models,'' \textit{J. Amer. Statist. Assoc.}, vol. 103, pp. 1534--1544, 2008.




\bibitem{ref28}
A. K. Rao, L. Muratori, E. D. Louis, C. B. Moskowitz, and K. S. Marder, ``Spectrum of gait impairments in presymptomatic and symptomatic Huntington's disease,'' \textit{Mov. Disord.}, vol. 23, pp. 1100--1107, 2008.

\bibitem{ref27}
J. O. Ramsay and B. W. Silverman, ``Functional data analysis,'' \textit{Encyclopedia Stat. Sci.}, vol. 4, pp. 1--15, 2004.



\bibitem{ref30}
Y. She, J. Shen, and C. Zhang, ``Supervised multivariate learning with simultaneous feature auto-grouping and dimension reduction,'' \textit{J. R. Stat. Soc. Ser. B}, vol. 84, pp. 912--932, 2022.

\bibitem{ref31}
X. Shen and H.-C. Huang, ``Grouping pursuit through a regularization solution surface,'' \textit{J. Amer. Statist. Assoc.}, vol. 105, pp. 727--739, 2010.

\bibitem{ref32}
D. Spiegelhalter, ``How old are you, really? Communicating chronic risk through effective age,'' \textit{BMC Med. Inform. Decis. Mak.}, vol. 16, pp. 1--6, 2016.

\bibitem{ref53}
L. Tang and P. X.-K. Song, ``Fused lasso approach in regression coefficients clustering: Learning parameter heterogeneity in data integration,'' \textit{J. Mach. Learn. Res.}, vol. 17, pp. 1--23, 2016.

\bibitem{ref54}
R. J. Tibshirani and J. Taylor, ``The solution path of the generalized lasso,'' \textit{Ann. Statist.}, vol. 39, pp. 1335--1371, 2011.


\bibitem{ref33}
B. Wang, X. Luo, Y. Zhao, and B. Caffo, ``Semiparametric partial common principal component analysis for covariance matrices,'' \textit{Biometrics}, vol. 75, pp. 1175--1186, 2019.

\bibitem{ref34}
Q.-S. Xu and Y.-Z. Liang, ``Monte Carlo cross validation,'' \textit{Chemom. Intell. Lab. Syst.}, vol. 56, pp. 1--11, 2001.



\bibitem{ref36}
C.-H. Zhang, ``Nearly unbiased variable selection under minimax concave penalty,'' \textit{Ann. Statist.}, vol. 38, pp. 894--942, 2010.

\bibitem{ref37}
Y. Zhang, R. Li, and C.-L. Tsai, ``Regularization parameter selection via generalized information criterion,'' \textit{J. Amer. Statist. Assoc.}, vol. 105, pp. 312--323, 2010.

\bibitem{ref38}
J. Zhou, N.-Y. Wang, and N. Wang, ``Functional linear model with zero-value coefficient function at sub-regions,'' \textit{Statistica Sinica}, vol. 23, pp. 25--50, 2013.



\end{thebibliography}
\end{document}